# Band Structure Modulation of $ZrO_2$ Nanoparticles for Control of CO Adsorption Properties: A Combined Density Functional Theory – Density Functional Tight Binding Study

Kexin Chen,[a] William Dawson,[b] Aulia Sukma Hutama,[c] Takahito Nakajima,[b] Keisuke Kameda,[a] Manabu Ihara,[a,1] Sergei Manzhos [a,2]

[a] School of Materials and Chemical Technology, Institute of Science Tokyo, Ookayama 2-12-1, Meguro-ku, Tokyo 152-8552, Japan

[b] RIKEN Center for Computational Science, Kobe, Hyogo 650-0047, Japan

[c] Department of Chemistry, Faculty of Mathematics and Natural Sciences, Universitas Gadjah Mada, Sekip Utara, Bulaksumur, Yogyakarta, 55281, Indonesia

## Abstract

We present a combined density functional theory (DFT) and density functional tight binding (DFTB) study of zirconia ($ZrO_2$) nanoparticles of experimentally relevant sizes of several nanometers and their interactions with the CO molecule. A hybrid DFTB–Force Field (DFTB–FF) framework is developed, whereby band structure calculations rely on an existing Slater–Koster framework, while the accuracy of structural optimization and adsorption properties is controlled by the introduction of classical long-range interatomic potentials into DFTB instead of the traditional repulsive potentials. Additionally, coordination-dependent Zr–C potentials are introduced to account for the distinct local chemical environments of bulk-like facet sites and under-coordinated tip and edge sites, thereby improving the description of CO adsorption. This hybrid DFTB–FF approach substantially improves the robustness of geometry optimization and provides a practical strategy for extending the applicability of DFTB to complex oxide nanostructures. The calculations reveal that termination stoichiometry can be used to engineer intrinsic, *p*-type, or *n*-type electronic structures and thereby tune the adsorption activity of zirconia nanoparticles. While stoichiometric nanoparticles do not activate the C–O bond, low-coordinated sites in Zr-rich (*n*-type) nanoparticles exhibit chemisorption accompanied by charge donation into a CO antibonding LUMO-derived orbital, resulting in C–O bond activation. These results demonstrate that stoichiometry-controlled electronic structure and under-coordinated surface sites introduced by nanostructuring play a key role in governing the adsorption strength and reactivity of zirconia nanoparticles.

[1] E-mail: mihara@chemeng.titech.ac.jp
[2] E-mail: manzhos.s.ss@m.titech.ac.jp

## 1 Introduction

Understanding and tuning heterogeneous interactions between gas-phase molecules and solid surfaces are central to the design of many advanced technologies, including gas sensing,[1,2] catalysis,[3,4] and energy conversion devices.[5–7] In particular, oxide-based materials, widely used as active surfaces or supports, play a crucial role in heterogeneous electrochemical technologies such as fuel cells, solar cells, and photocatalysts.[8–10] These technologies rely not only on the bulk properties of oxides, such as electron or ion transport and the optical band gap, but critically on surface-specific phenomena that govern adsorption, charge transfer, and reaction mechanisms at the interface.[8,11–13]

Zirconia is a versatile oxide material that has been extensively employed as a catalyst and catalyst support in heterogeneous catalysis, including direct synthesis of dimethyl carbonate from methanol and $CO_2$,[14] photocatalytic $CO_2$ reduction,[15] electrochemical reactions for fuel conversion in solid oxide cells,[16,17] and as a support for metals in processes such as methanol synthesis, methanol steam reforming, methanol decomposition, water–gas shift reaction, and dehydrogenative coupling reactions.[18–20] To better understand and design such systems, it is essential to probe the surface chemistry of zirconia, and to access the electronic structure-mediated mechanism of its interaction with gas-phase reactants and intermediates. In this work, we consider CO as a key adsorbent, as zirconia–CO interactions are of importance in a range of applications, such as isosynthesis,[21] methanol synthesis,[22] water-gas shift,[23,24] and reverse Boudouard reactions.[17,25]

Pure zirconium dioxide undergoes a phase transformation from monoclinic (stable at room temperature) to tetragonal (at about 1173 °C) and then to cubic (at about 2370 °C).[26] With the addition of yttria to pure zirconia, the replacement of $Zr^{4+}$ by $Y^{3+}$ ions on the cationic sublattice results in the stabilization of the cubic polymorph of zirconia over a wider range of temperatures.[26] While Y doping introduces oxygen vacancies and modifies local electronic properties, the fundamental surface chemistry and structure framework of YSZ remains closely related to that of pure $ZrO_2$.[27] Carbon monoxide serves as a well-established probe molecule to characterize metal oxide surfaces, owing to its well-understood electronic structure and sensitivity to surface charge, coordination, and acidity. By analysing CO adsorption behaviour such as adsorption energy, binding configuration, and charge transfer, one can gain insight into the electronic and chemical environment of zirconia surfaces and interaction mechanisms. Such systems can serve as a theoretical template model for exploring how structural features and surface electronic states influence molecule–surface interactions and for understanding more complex doped systems and the interaction with more complex carbon-containing adsorbates.

To date, most previous investigations have focused on periodic $ZrO_2$ surface models.[27–29] However, these flat surface models cannot adequately capture the complex morphology of real surfaces, particularly in terms of curvature effects, undercoordinated sites, and surface reconstructions. Additionally, materials at the nano scale exhibit special chemical, optical, electronic, and magnetic properties that differ markedly from their bulk counterparts.[30] The nanostructure's low dimensionality, high surface-to-volume ratio, and under-coordination of surface atoms suggest a possible strategy of chemical environment and size-dependent surface property modulation. The effects of molecules interacting with $ZrO_2$ nanoparticle (NP) surfaces have not been systematically addressed yet from an electronic structure-based perspective. To bridge this gap, this study aims at investigating CO adsorption on $ZrO_2$ nanoparticles, varying in size from a few hundred to thousands of atoms (approaching experimentally accessible scales),[31–33] to provide theoretical insights about nanosizing effects on the mechanism of CO molecule adsorption. We focus specifically on band structure modulation as a strategy to control molecule–surface interactions. Nanoparticles offer a unique way of band structure modulation by structure and stoichiometry, with a possibility of achieving intrinsic, *n*-type or *p*-type band structures without recourse to doping by heterogeneous atoms.

Nanoparticles pose a particular challenge for electronic structure methods, as systems of experimentally relevant sizes constitute intrinsically large-scale problems that make *ab initio* calculations computationally expensive. Computational reports to date on oxide nanoparticles, and in particular on zirconia nanoparticles, have focused on density functional theory (DFT) approaches.[34] The scaling of the computational cost of DFT with system size is formally cubic, limiting the exploration scale up to 700 atoms (about 3 nm in size) in previous reports.[35,36] While recent developments in linear-scaling DFT have enabled the investigation of nanoparticles beyond the size range accessible to conventional DFT,[37] these approaches remain computationally demanding for systematic studies involving molecular adsorption and extensive structural sampling. Consequently, DFT studies of molecular adsorption on zirconia nanoparticles are scarce, with only few works available.[38–40]

Self-consistent-charge Density Functional Tight Binding[41–43] is a promising alternative for the modelling of nanos-sized systems,[44–46] which offers a balance between accuracy and computational efficiency. DFTB is an approximate (semiempirical) method, in which the total energy is expressed, similar to DFT, as a functional of electron density, specifically, as a (up to the) third-order Taylor expansion around a reference (promolecular) electron density, $E[\rho(\boldsymbol{r})] = (E_0[\rho_0] + E_1[\rho_0, \delta\rho]) + E_2[\rho_0, (\delta\rho)^2] + E_3[\rho_0, (\delta\rho)^3]$. In the context of this work, the DFTB energy can be decomposed as

$$E^{DFTB} = E_{Band} + E_{SCC} + E_{Rep.} \tag{1}$$

Here, the band structure energy $E_{Band} = E_1$ is the sum of the orbital energies over all occupied orbitals. The self-consistent charge (SCC) term $E_{SCC} = E_2 + E_3$ accounts for the contributions arising from electron density changes due to bonding. Parameters for the Hamiltonian matrix elements and overlap integrals used to compute $E_{Band}$ are tabulated in Slater-Koster tables. The sum of $E_{Band}$ and $E_{SCC}$ is commonly referred to as the electronic energy in DFTB. The repulsive potential $E_{Rep} = E_0$ is represented by distance-dependent pairwise interactions that account primarily for short-range core–core and Pauli repulsion. In practice, it also serves as an empirical correction term to compensate for the approximations introduced in the electronic part of DFTB, enabling the method to reproduce reference energies, forces, and geometries obtained from higher-level theory calculations.[47,48]

Since the electronic terms are largely parameterized in a non-empirical manner, it is common practice to first determine electronic parameters and subsequently fit $E_{Rep}$ to a set of reference (training) systems to minimize the difference between DFTB and DFT energies, forces, and geometries.[49,50] A wide range of parameterization strategies has been proposed for repulsive potentials, including predefined analytical functional forms,[51] machine-learning-based approaches,[49,52] and various hybrid schemes in between.[53] However, although robust and transferable workflows have been established for the electronic part of DFTB, the parameterization of $E_{Rep}$ remains one of the most labour-intensive and system-dependent aspects of DFTB parameter development.[49] Instead of constructing an entirely new parameter set, this work explores an alternative strategy that extends the applicability of existing DFTB parameters by augmenting the original repulsive potentials with classical long-range force-field (FF) interactions. Within the proposed DFTB–FF framework, one separates the issues of the accuracy of the band structure and electronic properties governed by the Slater–Koster parameterization, e.g., band structure, density of states, charges and electronic state occupancy - quantities necessitating an electronic-structure-level treatment in the first place - from the issue of the accuracy of structures and interaction energies. The separation of electronic and structural descriptions allows the mature parameterization strategies developed in molecular mechanics to complement the DFTB framework without requiring reparameterization of the electronic terms. The proposed DFTB–FF approach provides a simple, transferable, and computationally efficient route for improving the geometric performance of existing DFTB parameter sets. It is enabled by the presence of the $E_{Rep}$ term in DFTB's governing equations, which in our approach is modified to include long-range potentials.

In this study, this approach is applied to study CO interactions with zirconia nanoparticles of different sizes with up to about 4,400 atoms (about 6 nm). The performance of the proposed method is benchmarked against DFT calculations for the structural optimization of zirconia nanoparticles, demonstrating improved geometric stability and acceptable accuracy. We further investigate the influence of local coordination environments on molecular adsorption using DFT and develop coordination-dependent Zr–C potentials for the DFTB–FF framework. This enables systematic exploration of size-dependent atomic and electronic structures beyond the practical limits of conventional DFT calculations, thereby helping to bridge the gap between theoretical simulations and experimentally relevant nanoparticle sizes. We demonstrate that controlling termination stoichiometry enables the tailoring of intrinsic, *p*-type, and *n*-type electronic structures, thereby modulating the catalytic activity of zirconia nanoparticles. Whereas stoichiometric nanoparticles show little activation of the C–O bond, under-coordinated sites in *n*-type nanoparticles promote chemisorption through electron donation into a CO antibonding LUMO-derived orbital, leading to C–O bond activation. These findings highlight the critical roles of nanostructure-modulated electronic structures and under-coordinated surface sites in determining the adsorption strength and reactivity of zirconia nanoparticles.

## 2 Methods

### *2.1 DFT calculations*

To assist DFTB parameterization and benchmarking, Kohn-Sham Density Functional Theory (KS-DFT) calculations were performed for periodic Zr–O–C systems to assess the bulk zirconia limit as well as for CO adsorption on $ZrO_2$ surfaces, using the Perdew-Burke-Ernzerhof (PBE) functional and Quantum ESPRESSO.[54,55] A kinetic energy cutoff of 40 Ry was applied for valence electrons, and the core electrons were represented using projector augmented wave (PAW) pseudopotentials. The convergence criterion for self-consistent field calculations was set to $1.0 \times 10^{-8}$ Ry, while the convergence thresholds for ionic optimization were set to $1.0 \times 10^{-4}$ Ry for the total energy and lower than $2.0\times10^{-2}$ eV/Å for the forces. Electronic occupancies were smeared using ordinary Gaussian spreading with a smearing width of $7 \times 10^{-3}$ Ry. Geometry optimizations of the nanoparticles were performed using only the Γ-point, with at least 10 Å of vacuum separating periodic images (a $35^3$ Å$^3$ vacuum box was used for the $Zr_{79}O_{160}$, $Zr_{80}O_{160}$, $Zr_{85}O_{160}$ nanoparticles, whereas a $45^3$ Å$^3$ vacuum box was used for the $Zr_{231}O_{448}$ nanoparticle and the corresponding adsorption calculations). For bulk and slab calculations, k-point meshes with a reciprocal-space density of at least 30 Å$^{-1}$ were used along the periodic directions. Spin-polarized calculations showed negligible effects on the structural and energetic properties relevant to the present study (although density minima resulting in small spin-polarized features in the PDOS were observed when initializing with spin). Therefore, all subsequent calculations were performed using the non-spin-polarized formalism to reduce the computational cost.

CO adsorption energies were computed as

$$E_{ad} = E_{CO/NP} - (E_{NP} + E_{CO}), \tag{2}$$

where $E_{CO}$, $E_{NP}$, and $E_{CO/NP}$ are the energies of the optimized (with the respective method) CO molecule, zirconia nanoparticle, and zirconia nanoparticle with the adsorbed CO molecule. During geometry optimization with the adsorbed CO molecule, the atoms in the outermost O–Zr–O layer were allowed to relax and bind to the CO molecule while the core atoms of the NPs were fixed to the geometries of the optimized NP to minimize contributions from minute structural reorganization of the nanoparticles to the calculated adsorption energy; small rearrangements can be a significant source of uncertainty due to the large number of atoms and absence of symmetry in doped NPs. After each NP–CO binding geometry was optimized, the CO molecule was removed, and the bare nanoparticle was reoptimized to determine whether the interaction with CO nudges the NP to a lower-energy structure. The nanoparticle energy obtained from this re-optimization was then used as $E_{NP}$ for calculating the adsorption energy.

### *2.2 DFTB calculations and parameterization*

Self-consistent charge DFTB[41,56] calculations were done in the DFTB3 regime using the DFTB+ code.[43,57,58] The convergence criterion for SCC calculation was set to $1.0\times10^{-6}$

|*e*|, while the ionic optimization convergence criteria were set to $5.0\times10^{-5}$ Hartree for energy and $1.0\times10^{-2}$ eV/Å for forces. An electronic temperature of 200 K was used to improve SCC convergence, using Fermi smearing.

*2.2.1 Modification of the Zr-Zr/Zr–O interactions with classical long-range force fields*

The standard DFTB3/3ob parameters for zirconia are tuned for biomedical and material science applications such as Zr-based metal-organic frameworks (e.g., UiO-66 and UiO-67),[48] but exhibit low accuracy in geometry optimization for zirconia solid oxides, particularly for NPs with complex coordination environments subjected to perturbations such as adsorption or thermal annealing that we used to assist optimization and test stability (see **Figure S1** and **Figure S2**), even though, as we will show later, the parameters provide a qualitatively correct partial density of states (PDOS) and electronic structure-based mechanism of CO–nanoparticle interaction. While the force field level of theory in principle allows the reproduction of structures and energetics, the electronic-structure-based mechanistic insight is the key advantage of DFTB. To enhance the stability of geometry optimization in DFTB, additional classical longer-range force fields (FF) were incorporated into the parameters via the repulsive potential input block for Zr–Zr/Zr–O/O–O atomic pairs, for which the added potentials were discretized on a grid and implemented as spline parameters. We use the recently proposed FF of Ref. [59] that consists of pairwise interatomic potentials (including Buckingham, Coulombic, Fermi and Gaussian terms, see Ref. [59] for details) developed to reproduce the structures and stability of $ZrO_2$ and YSZ polymorphs, and has been validated for grain boundary modeling with good consistency relative to *ab initio* results.[59] The introduction of the FFs leads to geometries more similar to those obtained from FF-based molecular dynamics optimizations, while retaining the DFTB (Slater-Koster) framework for electronic structure analysis that is valuable for understanding the mechanism of molecule–NP surface reactions. The contributions to the interatomic interactions are illustrated in **Figure 1 (a)**, where the pairwise potential part of the DFTB parameters for the zirconia system is described by a combination of the original DFTB repulsive potential and the added FF term.

The FF was introduced into the DFTB parameters using a weighting factor (see **Figure S3** for details). Hybrid DFTB–FF repulsive potential splines were generated by scaling the force-field contribution by factors ranging from 1 to 10 and interpolating the resulting potential using the CubicSpline function implemented in SciPy.[60] The cutoff radius for the splines was set to a long-range value of 30 Å to preserve the long-range Coulombic interactions inherent in the original force field models. To enable the application of the hybrid DFTB–FF framework to periodic systems (e.g., bulk/slab models used for benchmarking below) and nanoparticles with sizes larger than this cutoff, the Coulombic interaction in the original force field was modified using a screening function, allowing it to smoothly decay to zero at the cutoff radius:

$$E_{Coul/dsf} = q_i q_j \left[ \frac{erfc(\alpha r)}{r} - \frac{erfc(\alpha r_c)}{r_c} + \left( \frac{erfc(\alpha r_c)}{{r_c}^2} + \frac{2\alpha}{\sqrt{\pi}} \frac{\exp(-\alpha^2 {r_c}^2)}{r_c} \right)(r - r_c) \right], \quad (3)$$

where $erfc$ is the complementary error-function, and a damping parameter $\alpha$ of 0.05 and a cutoff $r_c$ of 30 Å are applied. The structural and energetic stability of DFTB–FF parameters was evaluated by DFTB-based molecular dynamics (MD) annealing simulations for the zirconia NP shown in **Figure S3**. The 3ob parameters with FF correction terms are able to handle violations of crystallinity and reproduce a reasonable description of periodic zirconia solids and NP structures (see **Table S1** and **Figure 3**). The Zr–Zr/Zr–O/O–O interatomic potential applied in this work is presented in **Figure S4**.

### *2.2.2 Parameterization of the Zr–C potential for zirconia nanoparticle–CO systems*

The parameterization of the Zr–C potential aims to accurately reproduce the adsorption behavior of CO on zirconia nanoparticles. Owing to the two-body pairwise nature of the DFTB repulsive potential, its transferability is strongly dependent on the local chemical environment. This limitation becomes particularly pronounced for clusters and nanoparticles with a large fraction of the surface atoms being under-coordinated.[47] To overcome this limitation, two distinct Zr–C repulsive potentials were developed: one for bulk-like coordinated facet sites and the other for under-coordinated tip and edge sites. The Zr atoms in these sites have significantly different charge states (see below). The use of environment-dependent interaction potentials is common in atomistic simulations.[61–63] The Zr–C potentials were parameterized using potential energy curves (PECs) calculated with DFT as reference. Within the DFTB3 formalism, the potentials were obtained by subtracting the DFTB electronic energy from the corresponding shifted DFT reference energy. Specifically, we utilized CO@$ZrO_2$(111) and CO@$Zr_{85}O_{160}$ systems with adsorption on the Zr facet sites and under-coordinated tip/edge sites of NPs as references (see **Figure S5**) for Zr–C potential fitting. The Zr–C bond lengths were systematically varied from 2.0 to 5.0 Å by placing the CO molecule relative to the adsorption site to generate the corresponding reference PECs with CO positioned at various adsorption sites on zirconia nanoparticles. The parameterization was performed using a semi-automated fitting code by introducing exponential and Gaussian correction terms to the original 3ob Zr–C repulsive potential. The cutoff radius of the Zr–C repulsive splines was extended to 30 Å, matching that adopted for the Zr–Zr/Zr–O interactions to ensure consistency within the long-range hybrid DFTB–FF framework. The parameters for the C–O, C–C, and O–O interactions were adopted directly from the 3ob-3-1 parameter set.[64]

To distinguish the parameter set for under-coordinated adsorption sites from that for bulk-like coordinated facet sites, the Zr–C potential for tip and edge sites was implemented using an additional atom type, denoted as $Zr_x$. Accordingly, the corresponding Slater–Koster files were labelled as $Zr_x$–C and C–$Zr_x$, whereas all other interactions involving $Zr_x$, including $Zr_x$–Zr, $Zr_x$–O, $Zr_x$–$O_x$ (where $O_x$ denotes the oxygen atom in the CO molecule to distinguish it from lattice oxygen atoms in zirconia), and their reverse pairs, were kept identical to those of the original Zr atom. The interatomic interactions for zirconia NP–CO systems with a Zr–$Zr_x$–O–C–$O_x$ parameter file regime are illustrated in **Figure 1 (b)**. During DFTB calculations, Zr atoms at under-coordinated tip and edge adsorption sites were assigned the $Zr_x$ atom type, thereby enabling the program to automatically read the appropriate Zr–C repulsive potential. The Zr–C repulsive potentials applied in this work are presented in **Figure S6**.

(a)

| Interatomic Potential | Zr | O |
|---|---|---|
| Zr | DFTB + FF | DFTB + FF |
| O | DFTB + FF | DFTB + FF |

(b)

| Interatomic Potential | Zr (full-coordinated site) | $Zr_x$ (under-coordinated site) | O | C | $O_x$ (O in CO) |
|---|---|---|---|---|---|
| Zr (full-coordinated site) | DFTB + FF | DFTB + FF | DFTB + FF | DFTB ($E_{rep}$ modified) | DFTB |
| $Zr_x$ (under-coordinated site) | DFTB + FF | DFTB + FF | DFTB + FF | DFTB ($E_{rep}$ modified) | DFTB |
| O | DFTB + FF | DFTB + FF | DFTB + FF | DFTB | DFTB |
| C | DFTB ($E_{rep}$ modified) | DFTB ($E_{rep}$ modified) | DFTB | DFTB | DFTB |
| $O_x$ (O in CO) | DFTB | DFTB | DFTB | DFTB | DFTB |

Figure 1. Schematic illustration of pairwise interatomic potentials in the hybrid DFTB–FF framework for (a) zirconia nanoparticles; (b) *n*-type zirconia nanoparticles binding to CO molecules. $Zr_x$ denotes a Zr atom at an under-coordinated tip/edge adsorption site, whereas $O_x$ denotes the oxygen atom in the CO molecule. These labels are introduced to distinguish them from bulk-like facet Zr atoms and lattice oxygen atoms in zirconia, respectively, for parameter assignment.

# 3 Results and discussion

## *3.1 Geometries and stoichiometry-modulated band structures of zirconia nanoparticles and benchmarking of DFTB–FF performance*

The nanoparticle structures were generated to approximate the Wulff construction,[65] which predicts the equilibrium shape of a crystalline particle by minimizing its overall surface energy. The Wulff construction built according to known surface energies for lowest-energy surfaces (surface energies of 1.20, 1.70 and 2.67 J $m^{-2}$ calculated for cubic $ZrO_2$ (111), (110), and (100)[66]) results in an octahedral shape exposing only (111) surfaces. The preferred termination of the (111) surface is O-terminated.[66] The NP models were therefore prepared by cleaving bulk c-$ZrO_2$ (cubic $ZrO_2$) along the most stable (111) surfaces, resulting in O-terminated octahedral nanoparticles shown in **Figure 2**. Similar modelling strategies have been used for $CeO_2$,[67,68] tetragonal-$ZrO_2$,[38] and $TiO_2$ nanoparticles.[69] This results in non-stoichiometric, Zr-excess structures (with compositions shown in **Figure 2**).

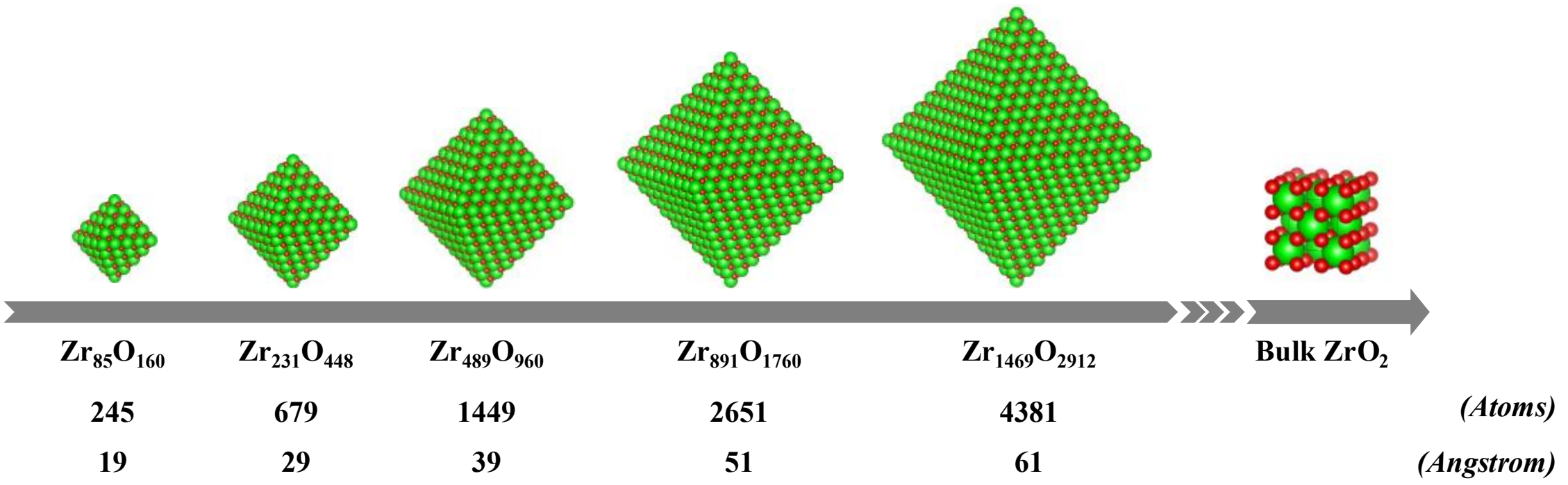

Figure 2. Models of octahedral zirconia nanoparticles and bulk $ZrO_2$. Atom color scheme used here and throughout unless otherwise stated: Zr, green; O, red. The numbers of atoms and nanoparticle sizes (defined as the maximum interatomic distances) are indicated below.

To modify the stoichiometry of zirconia NPs at the smallest scale shown in **Figure 2**, five or six Zr atoms were removed from the corners of the octahedral $Zr_{85}O_{160}$ NP. This corner truncation exposed local (100) facets and generated the stoichiometric $Zr_{80}O_{160}$ NP and O-excess $Zr_{79}O_{160}$ NP. **Figure 3** displays the geometries of the nanoparticles constructed using this approach and compares the structures optimized using DFT and DFTB–FF in stick models. Overall, the DFTB–FF optimized geometries show good agreement with their DFT counterparts. The band structures of zirconia nanoparticles, as a function of stoichiometry affecting O content, were investigated using DFTB–FF and validated with DFT calculations.

The PDOS for $Zr_{79}O_{160}$, $Zr_{80}O_{160}$, $Zr_{85}O_{160}$ nanoparticles optimized with DFT as well as the PDOS of these NPs optimized by DFTB are presented in **Figure 4**. The results indicate that the Zr:O stoichiometry modulates the band structure, with O-deficient, O-rich, and stoichiometric conditions exhibiting *n*-type, *p*-type, and intrinsic semi-conducting behaviour, respectively, as expected. The O-deficient $Zr_{85}O_{160}$ nanoparticle has an excess of electrons that are not compensated by O, introducing additional occupied electronic states near the bottom of the conduction band corresponding to Zr 4d orbital occupation. In contrast, the O-rich $Zr_{79}O_{160}$ nanoparticle creates extra unoccupied states near the valence band maximum, shifting the band structure towards *p*-type. The same modulation effect of the band structure by stoichiometry is observed in periodic bulk as well (**Figure S7)**. The *n*-doped states in the $Zr_{85}O_{160}$ nanoparticle suggest enhanced electron donation ability to adsorbate molecules interacting with zirconia and thereby enhanced surface activity.

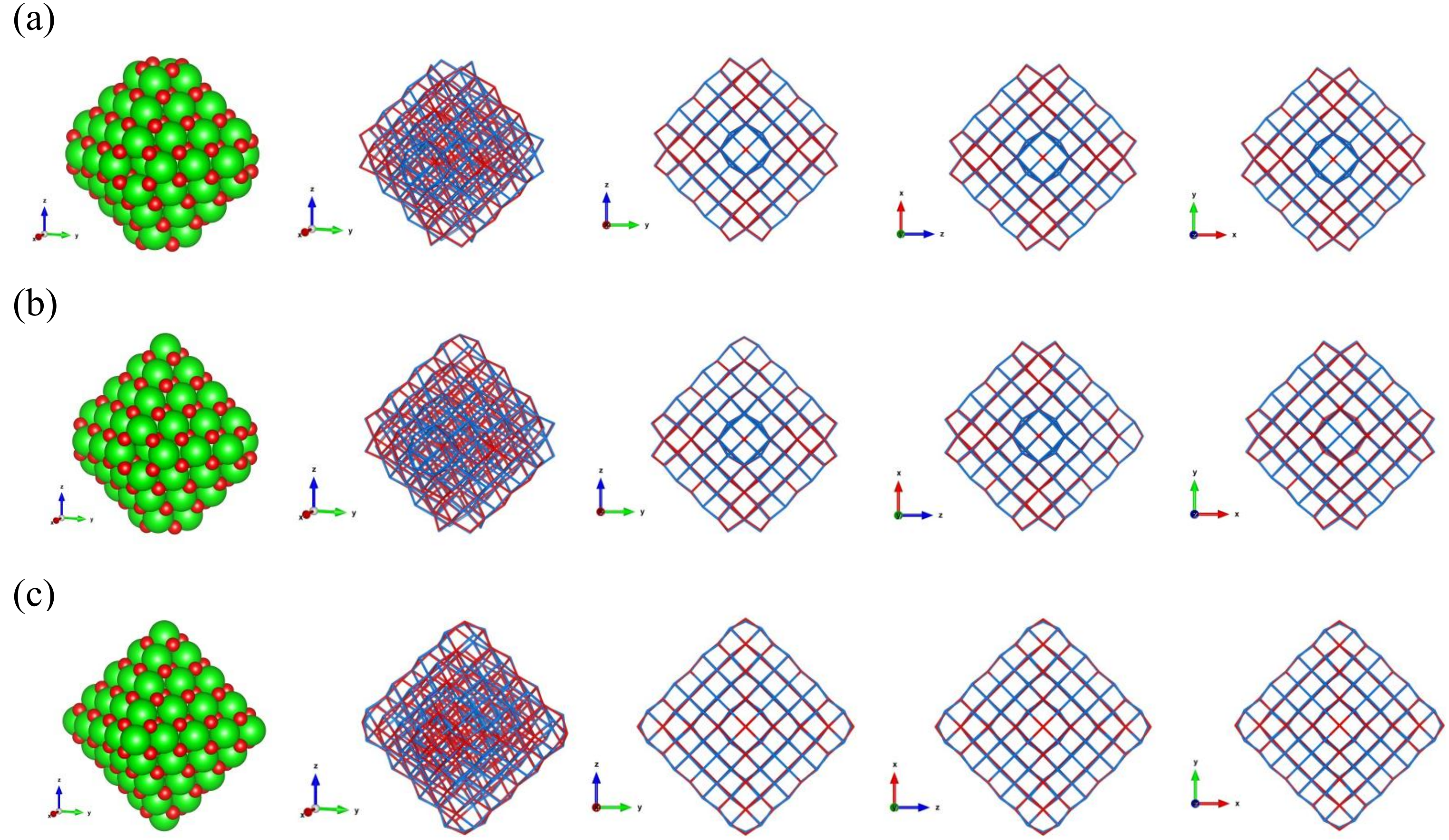

Figure 3. Optimized geometries of zirconia nanoparticles of (a) $Zr_{79}O_{160}$, (b) $Zr_{80}O_{160}$, (c) $Zr_{85}O_{160}$, presented as space-filling models showing the atomic positions, together with overlays of DFT (blue stick) and DFTB–FF (red stick) structures in different visualization views. The root mean square deviations (RMSD) of the atomic positions are calculated as 0.0716 Å, 0.0753 Å and 0.1026 Å, respectively.

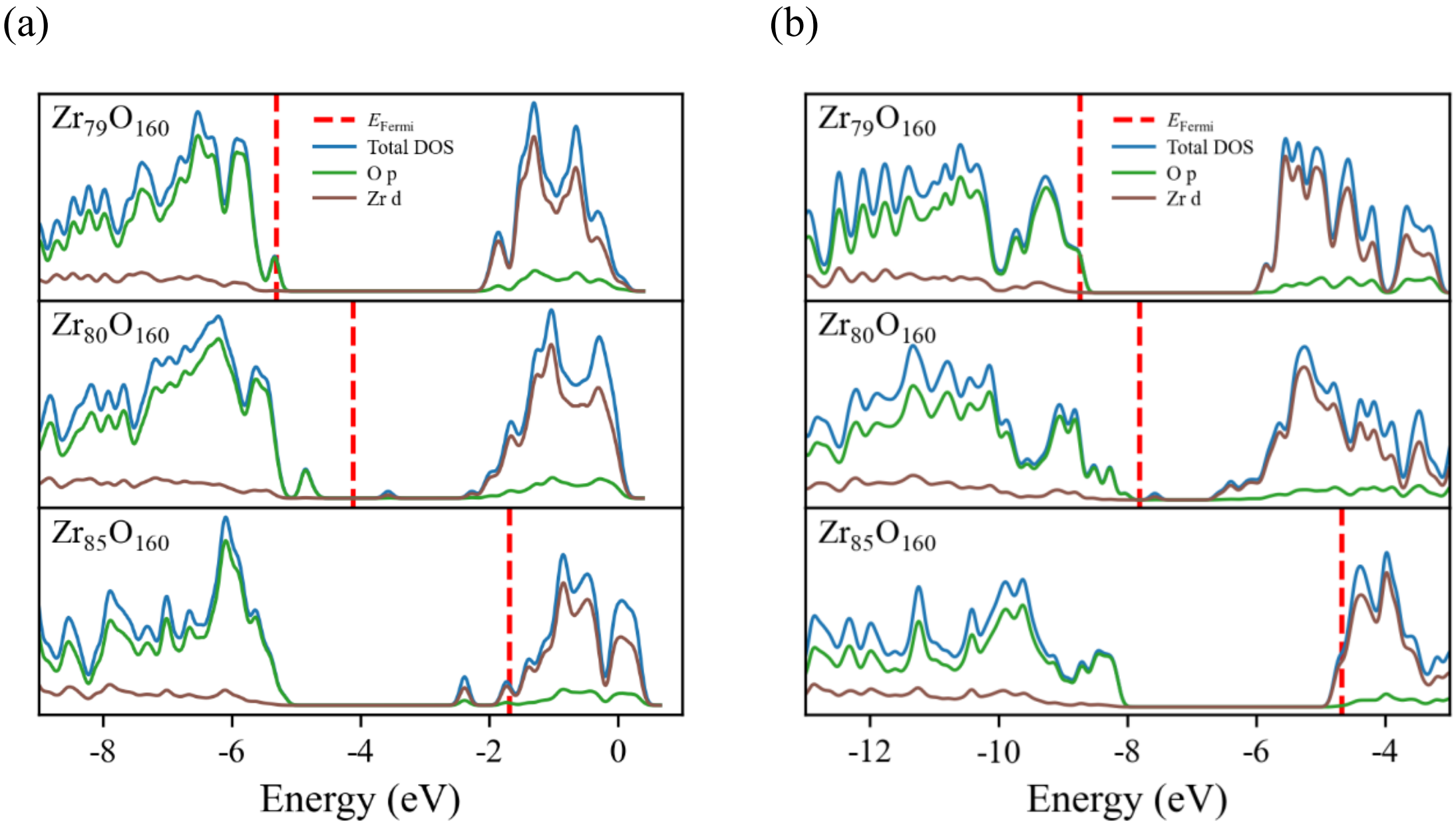

Figure 4. PDOS of $Zr_{79}O_{160}$, $Zr_{80}O_{160}$ and $Zr_{85}O_{160}$ computed with (a) DFT, (b) DFTB–FF.

Full optimization of zirconia nanoparticle and bulk models was conducted with the DFTB3 method to investigate the size effects. Due to the high computational cost, DFT geometry optimizations were carried out only for zirconia nanoparticles with the two smallest scales ($Zr_{85}O_{160}$, $Zr_{231}O_{448}$) to validate the DFTB–FF method. The relaxed NP structures exhibit a contraction in the average Zr–O bond lengths relative to bulk $ZrO_2$, as shown in **Table 1**. The averaged bond length of the nanoparticles increases as the size grows from ~250 atoms to 4,400 atoms, gradually approaching the bulk value, suggesting that shorter (stronger) bonds form at low-coordinated sites on nanoparticle surfaces. A similar lattice contraction effect in nanoclusters has been reported in both computational and experimental studies.[38,70,71]

Table 1. Optimized average Zr–O bond length (in Å) of octahedral zirconia nanoparticles and bulk computed with the DFTB–FF method. Values in parentheses correspond to DFT values.

| | $Zr_{85}O_{160}$ | $Zr_{231}O_{448}$ | $Zr_{489}O_{960}$ | $Zr_{891}O_{1760}$ | $Zr_{1469}O_{2912}$ | Bulk c-$ZrO_2$ |
|---|---|---|---|---|---|---|
| $d_{Zr-O}$ (Å) | 2.1627 (2.1769) | 2.1746 (2.1880) | 2.1808 | 2.1841 | 2.1996 | 2.2020 (2.2025) |

### *3.2 CO adsorption on zirconia nanoparticles and benchmarking of DFTB–FF performance*

CO adsorption properties of $Zr_{79}O_{160}$, $Zr_{80}O_{160}$, $Zr_{85}O_{160}$ nanoparticles with *p*-, intrinsic, and *n*-type band structures were explored with DFT. Multiple possible initial adsorption geometries on different sites were generated (**Figure S8**) on pre-optimized NP structures. The optimized structures revealed that stable adsorption occurs atop Zr atoms with the CO molecule coordinated to Zr via its C atom, with CO–nanoparticle distances ranging from approximately 2.3 to 2.6 Å. In contrast, the O atoms on the surface of the NP exhibited weak interactions with CO, repelling the molecule to distances exceeding 3.7 Å. These findings align with previous adsorption studies on zirconia surface and nanoparticle models.[27,39,40] The adsorption energy, C–O bond length, and charge transfer based on Bader analysis[72–76] to CO for Zr sites of $Zr_{79}O_{160}$, $Zr_{80}O_{160}$, $Zr_{85}O_{160}$ nanoparticles were summarized to characterize the adsorption performance (**Figure 5)**. C–O bond weakening together with bond length elongation and charge transfer from the nanoparticle to CO are expected in the case of chemisorption, which is desirable for promoting CO decomposition.

**Figure 5** shows adsorption energies, C–O bond lengths, and charge transfer to CO for different Zr adsorption sites on intrinsic, *p*-, and *n*-doped particles for the case of $Zr_{80}O_{160}$, $Zr_{79}O_{160}$, $Zr_{85}O_{160}$ nanoparticles calculated with DFT. The adsorption properties of different configurations (shown in **Figure 6**) are summarized in **Table 2**. The following is observed: In general, the under-coordinated Zr sites ($Zr_{tip}$, $Zr_{corner}$, $Zr_{edge}$) show a stronger adsorption strength than full-coordinated Zr sites ($Zr_{facet}$) for nanoparticles with different stoichiometries. The strongest $E_{ad}$ for $Zr_{79}O_{160}$, $Zr_{80}O_{160}$, $Zr_{85}O_{160}$ nanoparticles were calculated as -0.47 eV, -0.75 eV and -0.63 eV, respectively. The effect of a posteriori dispersion correction[77–79] on adsorption was also considered and was not substantial for the purpose of this work (see **Table S3)**. Although these nanoparticles exhibit comparable adsorption energies, the adsorption mechanisms differ significantly owing to their distinct electronic structures. When CO adsorbs on the tip site of $Zr_{85}O_{160}$, there is noticeable bond elongation up to 1.17 Å (elongation of 2.56%); by contrast, on the intrinsic and *p*-type nanoparticles the bond length ranges from 1.134 to 1.139 Å, which is close to the DFT-optimized bond length of free CO (1.141 Å). The C–O bond elongation indicates bond weakening correlating with charge donation from the nanoparticle to the antibonding LUMO-derived orbital (see **Figure 7**, to be explained in the next paragraph). Moreover, the data on charge donation **(Figure 5, Table 2**) with Bader analysis confirm noticeable charge transfer from the *n*-type NP to the CO molecule, ranging from -0.18 to -0.42 $|e|$ depending on the Zr adsorption sites, whereas no significant charge transfer was observed for CO binding to the intrinsic or *p*-type NPs. It is worth noting that relatively large adsorption energies do not always correlate with significant charge transfer to CO. For instance, adsorption energies of -0.75 eV for the $Zr_{tip}$ site of $Zr_{80}O_{160}$, and -0.63 eV for the $Zr_{edge2}$ site of $Zr_{85}O_{160}$ are observed without a corresponding (significant) increase in electron transfer. A previous theoretical study also reported large CO adsorption energies for a tetragonal-$Zr_{40}O_{80}$ nanoparticle, together with blue shifts in the CO stretching frequency,[40] indicating that a large adsorption energy alone does not necessarily result in efficient C–O bond activation. This observation suggests that under-coordinated sites can strengthen adsorption through geometric effects, whereas activation of the CO molecule and weakening of the C–O bond is predominantly controlled by the electronic structure of the nanoparticle, particularly the *n*-type band structure that promotes electron donation into CO antibonding states.

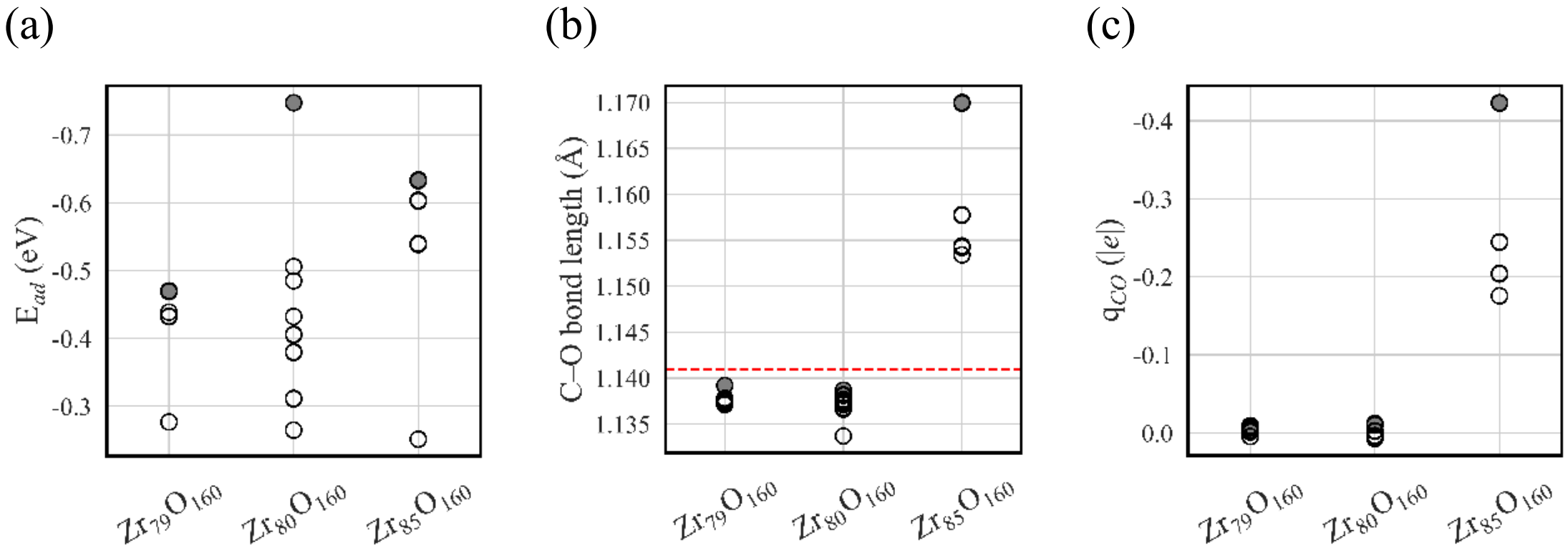

Figure 5. CO adsorption properties of $Zr_{79}O_{160}$, $Zr_{80}O_{160}$, and $Zr_{85}O_{160}$ nanoparticles at different sites. (a) CO adsorption energy; (b) C–O bond length, with the red dashed line indicating the bond length of a free CO molecule; (c) Bader charge on adsorbed CO. (Black hollow circles: data points of different adsorption sites; Gray solid dots: data points with the strongest adsorption properties.)

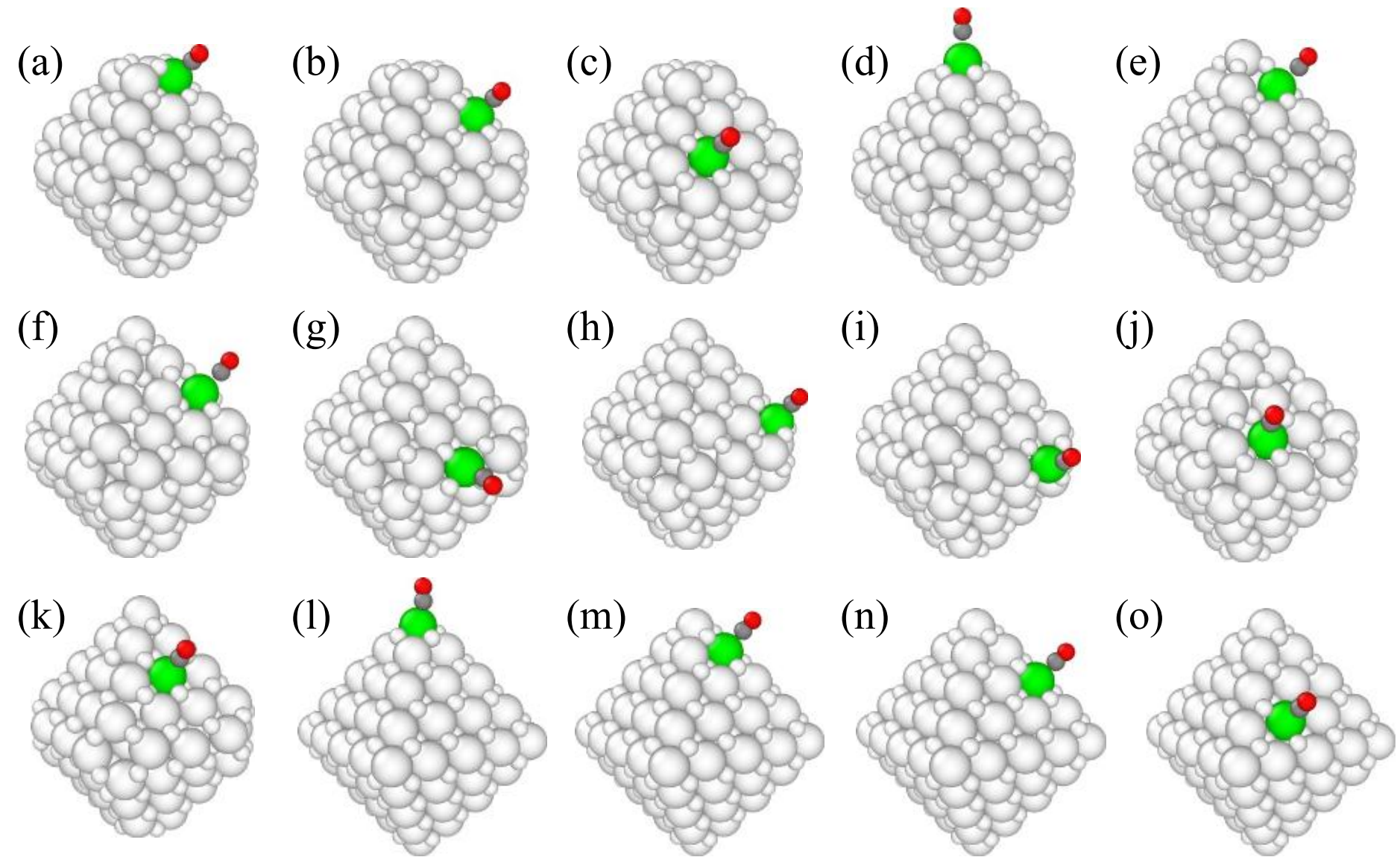

Figure 6. The CO adsorption configurations on different Zr adsorption atom sites of (a-c) $Zr_{79}O_{160}$, (d-k) $Zr_{80}O_{160}$, and (l-o) $Zr_{85}O_{160}$ nanoparticles optimized with DFT. Zr adsorption sites, C and O atoms are colored in green, dark grey and red, respectively, while other atoms in nanoparticles are colored in white for a visually clear contrast.

Table 2. CO adsorption energy ($E_{ad}$, eV), bond length of adsorbed CO ($d_{C-O}$, Å), adsorbed CO charge population ($q_{CO}$, $|e|$), and distance from the C atom to the nanoparticle ($d_{Zr-C}$, Å) at different Zr adsorption sites of $Zr_{79}O_{160}$, $Zr_{80}O_{160}$, and $Zr_{85}O_{160}$ nanoparticles computed with DFT.

| System | Type of band structure | Fig. 6 | Adsorption site | $E_{ad}$ | $d_{C-O}$ | $q_{CO}$ | $d_{Zr-C}$ |
|---|---|---|---|---|---|---|---|
| $Zr_{79}O_{160}$ | *p* | (a) | $Zr_{corner}$ | -0.47 | 1.138 | -0.01 | 2.51 |
| | | (b) | $Zr_{edge}$ | -0.43 | 1.137 | 0.00 | 2.55 |
| | | (c) | $Zr_{facet}$ | -0.28 | 1.139 | 0.01 | 2.49 |
| $Zr_{80}O_{160}$ | *intrinsic* | (d) | $Zr_{tip}$ | -0.75 | 1.134 | 0.01 | 2.54 |
| | | (e) | $Zr_{edge1}$ | -0.31 | 1.137 | 0.00 | 2.56 |
| | | (f) | $Zr_{edge2}$ | -0.41 | 1.137 | 0.01 | 2.57 |
| | | (g) | $Zr_{edge3}$ | -0.43 | 1.137 | 0.01 | 2.57 |
| | | (h) | $Zr_{corner1}$ | -0.51 | 1.138 | -0.01 | 2.51 |
| | | (i) | $Zr_{corner2}$ | -0.49 | 1.137 | 0.00 | 2.53 |
| | | (j) | $Zr_{facet1}$ | -0.38 | 1.138 | -0.01 | 2.51 |
| | | (k) | $Zr_{facet2}$ | -0.26 | 1.139 | -0.01 | 2.51 |
| $Zr_{85}O_{160}$ | *n* | (l) | $Zr_{tip}$ | -0.60 | 1.170 | -0.42 | 2.32 |
| | | (m) | $Zr_{edge1}$ | -0.54 | 1.154 | -0.20 | 2.40 |
| | | (n) | $Zr_{edge2}$ | -0.63 | 1.158 | -0.24 | 2.38 |
| | | (o) | $Zr_{facet}$ | -0.25 | 1.153 | -0.18 | 2.38 |

* The bond length of a free CO molecule calculated with DFT is 1.141 Å.

The weakening of the C–O bond via charge donation into the antibonding LUMO-derived orbital is confirmed in **Figure 7**, where the population of this orbital is shown for the *n*-type nanoparticle. In this case, the more highly occupied Zr *d* states act as electron donors, donating charge into the orbital corresponding to the antibonding $2\pi^*$ orbital of the free CO molecule. Therefore, adsorption on the electron donor-rich Zr sites of the *n*-type nanoparticle results in stronger CO binding accompanied by significant elongation of the C–O bond. In contrast, for adsorption cases on intrinsic and *p*-type nanoparticles, no appreciable electron transfer from the nanoparticles to CO is observed (see **Figure S9**).

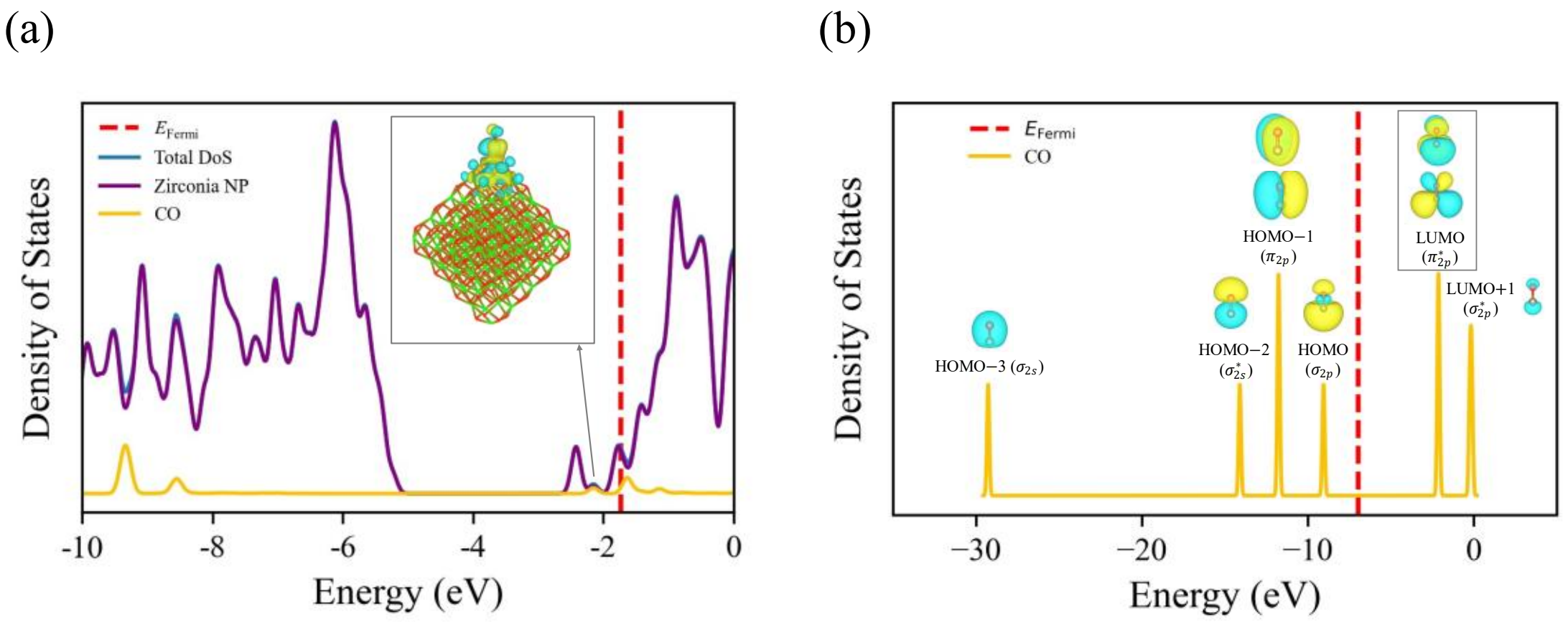

Figure 7. Densities of states computed with DFT for (a) CO adsorbed on $Zr_{85}O_{160}$, (b) free CO. The CO PDOS in (a) is scaled by a factor of three for clarity. The molecular orbital corresponding to the LUMO-derived orbital is shown in the inset of (a), while all molecular orbitals of the isolated CO molecule are shown in the inset of (b).

The extent of charge transfer and its influence on CO adsorption strength strongly depend on the electronic structure of the adsorption site, which is governed by the local chemical environment, including the nanoparticle stoichiometry and the coordination of the adsorption site. Particularly important is how the local coordination environment modulates the electronic state of Zr atoms in the nanoparticles. This can be glimpsed from Bader charges reported in **Table S2**, revealing that for $Zr_{85}O_{160}$ nanoparticle $Zr_{facet}$ sites carry a charge of approximately +2.56 |*e*|, comparable to that of bulk Zr atoms (+2.66 |*e*|). In contrast, under-coordinated surface sites are more electron-rich and exhibit lower positive charges, with values of +2.45/+2.51 |*e*| for $Zr_{edge}$ sites and a more reduced charge state of +1.70 |*e*| for $Zr_{tip}$ site, indicating an enhanced electron-donating capability. Since electron transfer from the nanoparticle into the antibonding orbitals of CO is the primary mechanism governing CO adsorption and C–O activation observed for *n*-type nanoparticles, the adsorption behavior is expected to be highly sensitive to the local electronic environment of the adsorption site. Two coordination-dependent Zr–C repulsive potentials were developed to reproduce the DFT adsorption behavior of $n$-type zirconia nanoparticles, with separate parameterizations for bulk-like facet sites, and for under-coordinated tip and edge sites within the DFTB–FF framework (see parameterization details in Supplementary Information).

Table 3. CO adsorption energy ($E_{ad}$, eV) and distance from the C atom to the nanoparticle ($d_{Zr-C}$, Å) at different Zr adsorption sites of $Zr_{85}O_{160}$ and $Zr_{231}O_{448}$ nanoparticles computed with DFT and DFTB–FF methods.

| System | Adsorption site | Fig. 8 | DFT | | DFTB–FF | |
|---|---|---|---|---|---|---|
| | | | $E_{ad}$ | $d_{Zr-C}$ | $E_{ad}$ | $d_{Zr-C}$ |
| $Zr_{85}O_{160}$ | $Zr_{tip}$ | (a) | -0.60 | 2.32 | -0.62 | 2.36 |
| | $Zr_{edge1}$ | (b) | -0.54 | 2.40 | -0.55 | 2.60 |
| | $Zr_{edge2}$ | (c) | -0.63 | 2.38 | -0.59 | 2.62 |
| | $Zr_{facet}$ | (d) | -0.25 | 2.38 | -0.28 | 2.38 |
| $Zr_{231}O_{448}$ | $Zr_{tip}$ | (e) | -0.63 | 2.30 | -0.62 | 2.37 |
| | $Zr_{edge1}$ | (f) | -0.58 | 2.38 | -0.57 | 2.63 |
| | $Zr_{edge2}$ | (g) | -0.67 | 2.35 | -0.61 | 2.64 |
| | $Zr_{edge3}$ | (h) | -0.71 | 2.35 | -0.61 | 2.64 |
| | $Zr_{facet1}$ | (i) | -0.33 | 2.35 | -0.24 | 2.41 |
| | $Zr_{facet2}$ | (j) | -0.31 | 2.34 | -0.16 | 2.46 |
| | $Zr_{facet3}$ | (k) | -0.39 | 2.33 | -0.21 | 2.42 |

**Table 3** summarizes CO adsorption properties of $n$-type $Zr_{85}O_{160}$ nanoparticles calculated using the hybrid DFTB–FF method and benchmarked against DFT results. The absolute differences in adsorption energies between the DFTB–FF and DFT calculations are within 0.05 eV, demonstrating good agreement in reproducing interaction energies between CO and the $Zr_{85}O_{160}$ nanoparticle. Although the DFTB–FF method slightly overestimates the adsorption distances, the optimized NP–CO binding geometries are highly consistent with those obtained from DFT as shown in **Figure 8**. The transferability of DFTB–FF is further tested on larger octahedral $n$-type $Zr_{231}O_{448}$ nanoparticle and is benchmarked against DFT results in **Table 3** and **Figure 8**, indicating acceptable accuracy in reproducing $E_{ad}$ and the geometries of various adsorption sites.

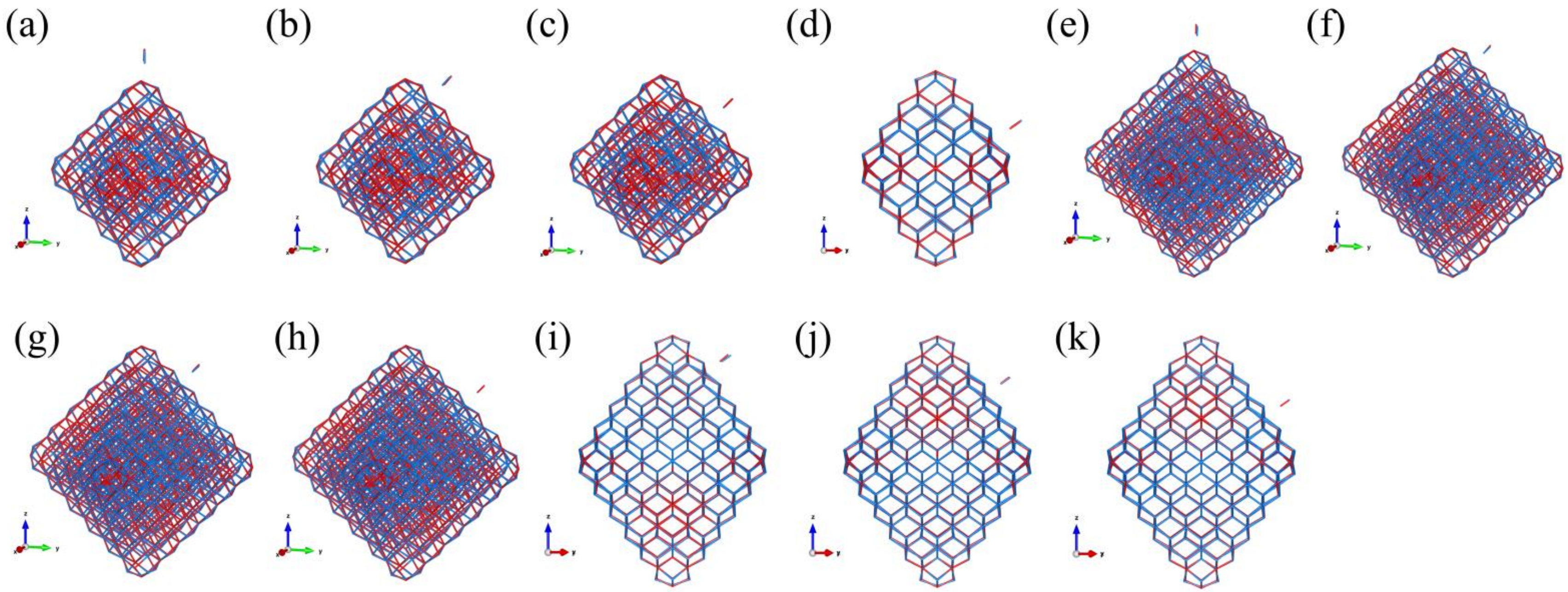

Figure 8. The CO adsorption configurations on different Zr adsorption atom sites of (a-d) $Zr_{85}O_{160}$ and (e-k) $Zr_{231}O_{448}$ nanoparticles, presented as an overlay of DFT (blue stick) and DFTB–FF (red stick) structures. The configurations of CO adsorbing on facet Zr atoms are shown in side views for clear visualization.

The strength and mechanism of CO molecule adsorption on zirconia nanoparticles were further investigated using DFTB–FF method for particle sizes increasing to $Zr_{489}O_{960}$, $Zr_{891}O_{1760}$ and as large as more than 4300 atoms in $Zr_{1469}O_{2912}$, which is beyond the size range accessible to DFT. As demonstrated above, the $Zr_{tip}$ sites exhibit pronounced adsorption activity with enhanced charge transfer to the adsorbed CO molecule. Moreover, $Zr_{tip}$ sites across nanoparticles of different sizes share a similar under-coordinated local environment, allowing them to serve as representative adsorption sites for evaluating size-dependent adsorption behavior. Therefore, for nanoparticles with scales beyond 1000 atoms, only the $Zr_{tip}$ sites were tested for CO adsorption calculations to manage computational cost. The adsorption properties, including the CO adsorption energy, C–O bond length, CO charge, and NP–CO distance, are summarized in **Table 4**, and the optimized configurations are shown in **Figure 9**. It should be noted that the charges reported by DFTB+ are Mulliken charges, which tend to underestimate the extent of total charge redistribution with limited accuracy.[80,81] To obtain a more reliable description of charge transfer, Bader charge analysis was performed by integrating the DFTB–FF valence-electron density within the atomic basins defined by a reconstructed all-electron density. The all-electron density was reconstructed by superimposing core-electron densities computed for isolated atoms using Gaussian 16 at the B3LYP/Sapporo-DZP-2012 level of theory onto the DFTB valence-electron density for all atoms in the system.

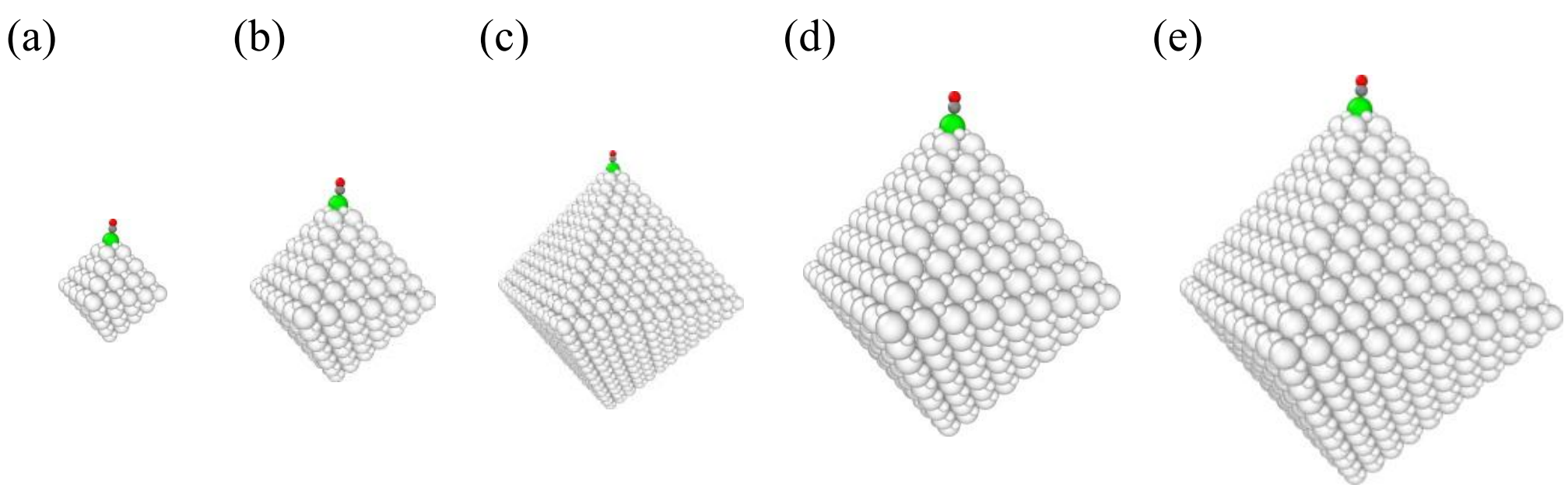

Figure 9. The CO adsorption configurations at the tip Zr sites of (a) $Zr_{85}O_{160}$, (b) $Zr_{231}O_{448}$, (c) $Zr_{489}O_{960}$, (d) $Zr_{891}O_{1760}$ and (e) $Zr_{1469}O_{2912}$ nanoparticles optimized with DFTB–FF. Zr adsorption sites, C and O atoms are colored in green, dark grey and red, respectively, while other atoms in nanoparticles are colored in white.

Table 4. CO adsorption energy ($E_{ad}$, eV), bond length of adsorbed CO ($d_{C-O}$, Å), adsorbed CO charge population ($q_{CO}$, |$e$|), and distance from the C atom to the nanoparticle ($d_{Zr-C}$, Å) on the tip Zr sites of $n$-type $Zr_{85}O_{160}$, $Zr_{231}O_{448}$, $Zr_{489}O_{960}$, $Zr_{891}O_{1760}$ and $Zr_{1469}O_{2912}$ nanoparticles.

| system | $E_{ad}$ | $d_{C-O}$ | $q_{CO}$ | $d_{Zr-C}$ |
|---|---|---|---|---|
| $Zr_{85}O_{160}$ | -0.62 | 1.136 | -0.14 | 2.36 |
| $Zr_{231}O_{448}$ | -0.62 | 1.130 | -0.09 | 2.37 |
| $Zr_{489}O_{960}$ | -0.64 | 1.125 | -0.06 | 2.38 |
| $Zr_{891}O_{1760}$ | -0.65 | 1.122 | -0.03 | 2.38 |
| $Zr_{1469}O_{2912}$ | -0.67 | 1.125 | -0.05 | 2.37 |

* The bond length of a free CO molecule calculated with DFTB3 is 1.104 Å.

The CO adsorption properties indicate a by and large similar adsorption energy for nanoparticles of different sizes up to a scale of approximately 4300 atoms. The fundamental adsorption mechanism, in which a state derived from the LUMO of the CO molecule becomes occupied via electron transfer, remains consistent across all nanoparticle sizes as shown in **Figure 10,** with CO's PDOS scaled by a multiple proportional to the total number of atoms in the system to achieve visible peak heights for ease of visualization. Moreover, charge donation into the CO 2π* orbital was also observed on these $n$-doped nanoparticles, indicating that the underlying interaction mechanism between CO and zirconia surfaces is qualitatively similar in that the interaction is primarily governed by an $n$-type electronic structure. As the NP size increases, there is a mild trend of weaker charge donation and smaller CO bond elongation. This corresponds to the LUMO-derived state in **Figure 10** becoming more localized on the NP and less on the CO molecule for larger systems.

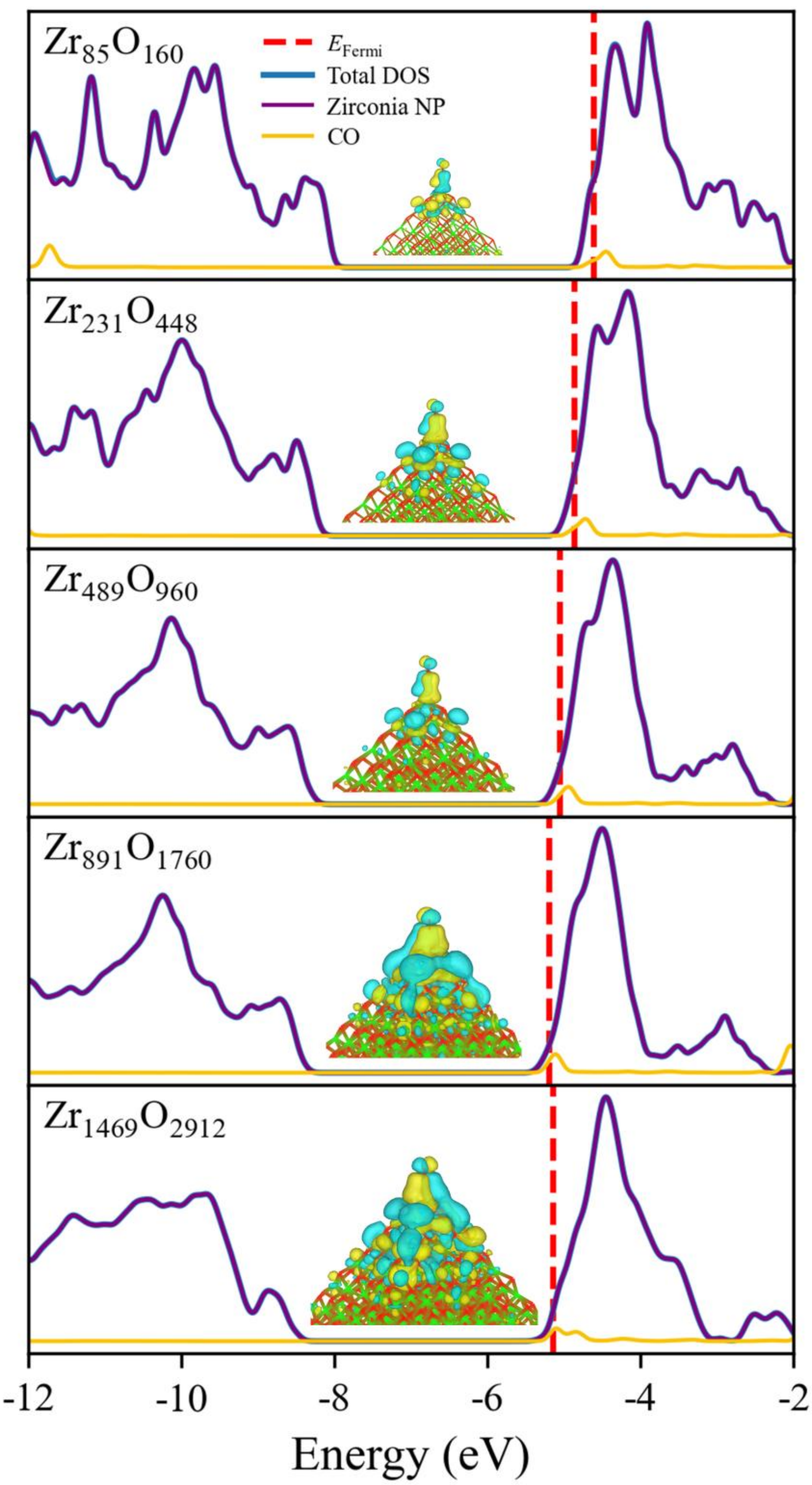

Figure 10. PDOS of CO adsorption on $Zr_{85}O_{160}$, $Zr_{231}O_{448}$, $Zr_{489}O_{960}$, $Zr_{891}O_{1760}$, and $Zr_{1469}O_{2912}$ nanoparticles. The corresponding LUMO-derived CO states around the HOMOs of the adsorption systems are shown as insets.

# 4 Conclusions

In this work, a comprehensive DFTB–DFT study was conducted to investigate the geometric, electronic structural, and molecular adsorption properties of nanoparticular cubic $ZrO_2$. The calculations employed a hybrid DFTB–FF approach, in which the original DFTB repulsive potentials were augmented with classical long-range force fields for the Zr–Zr, Zr–O, and O–O atomic pairs. This hybrid approach significantly improved the stability of geometry optimization by incorporating the structural robustness of the force field while preserving the DFTB framework for electronic-structure calculations. Furthermore, the CO adsorption properties of zirconia nanoparticles were systematically investigated using DFT, and coordination-dependent Zr–C repulsive potentials were developed to enable efficient nanoscale modeling of CO adsorption within the DFTB framework. Our approach partially decouples the accuracy of electronic properties, determined primarily by the Slater–Koster parameterization (e.g., band structure, density of states, atomic charges, and electronic-state occupancies), from that of structural and interaction-energy predictions, which can be improved through modification of the interatomic potentials. We achieve this within the DFTB framework itself thanks to the presence of two-body potential terms in the DFTB concept. This provides a practical strategy for improving the description of diverse chemical environments with DFTB by two-body interatomic potentials. This in turn facilitates the application of DFTB to systems that are computationally impractical with DFT, allowing us to harness the substantial computational cost advantage of DFTB. For example, our single-point benchmark for $Zr_{85}O_{160}$, performed on the Nibi cluster of the Digital Research Alliance of Canada using 96 MPI processes on compute nodes equipped with Intel 6972P processors (2.4 GHz), required 6153.2 s with DFT and only 12.9 s with DFTB, with comparable numbers of self-consistency iterations (55 and 44, respectively), corresponding to a nearly 500-fold speedup. To the best of our knowledge, this work represents the first DFTB study of zirconia nanoparticles for catalytic applications.

Our results demonstrate that modulation of the band structure via Zr:O stoichiometry serves as an effective strategy to control adsorption behavior, particularly through tuning the energy alignment of surface-localized electron donor states. $(ZrO_{2-x})_n$ nanoparticles with *n*-type electronic structures exhibited enhanced CO adsorption properties, characterized by elongation of the C–O bond and charge transfer into the CO 2π* antibonding orbital. In contrast, intrinsic $(ZrO_2)_n$ nanoparticles and *p*-type $(ZrO_{2+x})_n$ nanoparticles do not activate C-O bond even though they may show a comparable adsorption strength. Size-dependent studies across zirconia nanoparticles ranging from ~250 to ~4400 atoms further revealed that nano-structuring enhances CO activation due to the presence of low-coordinated and electronically active surface sites. Importantly, we have shown that although under-coordinated sites on nanoparticles or nanostructures can strengthen adsorption energy, weakening of the C–O bond is governed by an electronic-structure-mediated mechanism enabled by the *n*-type band structure.

Overall, this study underscores the importance of electronic structure engineering at the nanoscale and demonstrates the utility of incorporating long-range interatomic potentials into the DFTB framework for studies of nanosized systems, providing a strategy that may be applicable to other material systems beyond those considered here.

## 5 Author contributions

Kexin Chen: Calculations, Methodology, Data analysis, Writing - Original draft preparation, Writing - Reviewing and Editing. William Dawson: Calculations, Data analysis, Writing - Reviewing and Editing. Aulia Sukma Hutama: Calculations, Data analysis, Writing - Reviewing and Editing. Takahito Nakajima: Writing - Reviewing and Editing. Keisuke Kameda: Supervision, Writing - Reviewing and Editing. Manabu Ihara: Resources, Writing- Reviewing and Editing, Project administration, Supervision. Sergei Manzhos: Conceptualization, Methodology, Writing - Original draft preparation, Writing - Reviewing and Editing, Project administration, Supervision.

## 6 Conflicts of interest

There are no conflicts to declare.

## 7 Data availability

The data supporting this article and the Slater–Koster (SK) files used in this work have been included as part of the ESI.

## 8 Acknowledgements

This work was supported by JST Commercialization Support (Grant Number JPMJSF2514) and the Science Tokyo Academy of Energy and Informatics (ISE). We acknowledge computational resources provided through the TSUBAME Encouragement Program for Young/Female Users and by the Digital Research Alliance of Canada. We also thank Prof. Thomas S. Hofer and Armin Penz for providing the wavefunction coefficient files.

## 9 References

1 A. Dey, *Mater. Sci. Eng., B*, 2018, **229**, 206–217.
2 S. A. I. Steiner, T. F. Baumann, B. C. Bayer, R. Blume, M. A. Worsley, W. J. MoberlyChan, E. L. Shaw, R. Schlögl, A. J. Hart, S. Hofmann and B. L. Wardle, *J. Am. Chem. Soc.*, 2009, **131**, 12144–12154.
3 G. A. Somorjai and Y. Li, *Introduction to Surface Chemistry and Catalysis*, John Wiley & Sons, Hoboken, New Jersey, 2010.
4 G. Ertl, H. Knözinger, F. Schüth and J. Weitkamp, *Handbook of heterogeneous catalysis*, Wiley-VCH, Weinheim, Germany, 2008.
5 C. Feng, Z.-P. Wu, K.-W. Huang, J. Ye and H. Zhang, *Adv. Mater.*, 2022, **34**, 2200180.
6 Y. P. Zhu, C. Guo, Y. Zheng and S.-Z. Qiao, *Acc. Chem. Res.*, 2017, **50**, 915–923.
7 Q. Zhang, E. Uchaker, S. L. Candelaria and G. Cao, *Chem. Soc. Rev.*, 2013, **42**, 3127–3171.
8 V. E. Henrich and P. A. Cox, *The Surface Science of Metal Oxides*, Cambridge University Press, 1994.
9 M. S. S. Danish, A. Bhattacharya, D. Stepanova, A. Mikhaylov, M. L. Grilli, M. Khosravy and T. Senjyu, *Metals*, 2020, **10**, 1604.
10 X. Yu, T. J. Marks and A. Facchetti, *Nat. Mater.*, 2016, **15**, 383–396.
11 V. E. Henrich and P. A. Cox, *Appl. Surf. Sci.*, 1993, **72**, 277–284.
12 L. Wang, C. Shi, L. Wang, L. Pan, X. Zhang and J.-J. Zou, *Nanoscale*, 2020, **12**, 4790–4815.
13 A. V. Akimov, A. J. Neukirch and O. V. Prezhdo, *Chem. Rev.*, 2013, **113**, 4496–4565.
14 K. Tomishige, Y. Ikeda, T. Sakaihori and K. Fujimoto, *J. Catal.*, 2000, **192**, 355–362.
15 K. Sayama and H. Arakawa, *J. Phys. Chem.*, 1993, **97**, 531–533.
16 S. C. Singhal and K. Kendall, *High-temperature Solid Oxide Fuel Cells: Fundamentals, Design and Applications*, Elsevier, Amsterdam, Netherlands, 2003.
17 K. Kameda, S. Manzhos and M. Ihara, *J. Power Sources*, 2021, **516**, 230681.
18 N. Scotti, F. Bossola, F. Zaccheria and N. Ravasio, *Catalysts*, 2020, **10**, 168.
19 V. Idakiev, T. Tabakova, A. Naydenov, Z.-Y. Yuan and B.-L. Su, *Appl. Catal., B*, 2006, **63**, 178–186.
20 K. Li and J. G. Chen, *ACS Catal.*, 2019, **9**, 7840–7861.
21 Y. Li, D. He, Q. Zhu, X. Zhang and B. Xu, *J. Catal.*, 2004, **221**, 584–593.
22 M.-Y. He and J. G. Ekerdt, *J. Catal.*, 1984, **87**, 381–388.

23 E. Chenu, G. Jacobs, A. C. Crawford, R. A. Keogh, P. M. Patterson, D. E. Sparks and B. H. Davis, *Appl. Catal., B*, 2005, **59**, 45–56.
24 E. I. Kauppi, K. Honkala, A. O. I. Krause, J. M. Kanervo and L. Lefferts, *Top. Catal.*, 2016, **59**, 823–832.
25 M. Kogler, E.-M. Köck, B. Klötzer, T. Schachinger, W. Wallisch, R. Henn, C. W. Huck, C. Hejny and S. Penner, *J. Phys. Chem. C*, 2016, **120**, 1795–1807.
26 H. Yanagida, K. Koumoto and M. Miyayama, *The Chemistry of Ceramics*, Wiley, 1996.
27 A. Cadi-Essadek, A. Roldan and N. H. de Leeuw, *Surf. Sci.*, 2016, **653**, 153–162.
28 S. Jalili and M. Keshavarz, *Comput. Theor. Chem.*, 2020, **1173**, 112702.
29 T. Joutsuka and S. Tada, *J. Phys. Chem. C*, 2023, **127**, 6998–7008.
30 A. P. Alivisatos, *Science*, 1996, **271**, 933–937.
31 S. Chen, A. Tennakoon, K.-E. You, A. L. Paterson, R. Yappert, S. Alayoglu, L. Fang, X. Wu, T. Y. Zhao, M. P. Lapak, M. Saravanan, R. A. Hackler, Y.-Y. Wang, L. Qi, M. Delferro, T. Li, B. Lee, B. Peters, K. R. Poeppelmeier, S. C. Ammal, C. R. Bowers, F. A. Perras, A. Heyden, A. D. Sadow and W. Huang, *Nat. Catal.*, 2023, **6**, 161–173.
32 S. Raj, M. Hattori and M. Ozawa, *Mater. Lett.*, 2019, **234**, 205–207.
33 R. Sigwadi, T. Mokrani and M. Dhlamini, *Phys. B (Amsterdam, Neth.)*, 2020, **581**, 411842.
34 A. Fernando, K. L. D. M. Weerawardene, N. V. Karimova and C. M. Aikens, *Chem. Rev.*, 2015, **115**, 6112–6216.
35 E. Albanese, A. Ruiz Puigdollers and G. Pacchioni, *ACS Omega*, 2018, **3**, 5301–5307.
36 A. R. Puigdollers, F. Illas and G. Pacchioni, *J. Phys. Chem. C*, 2016, **120**, 4392–4402.
37 S. Li, T. Miyazaki and A. Nakata, *Phys. Chem. Chem. Phys.*, 2024, **26**, 20251–20260.
38 A. Ruiz Puigdollers, F. Illas and G. Pacchioni, *Rend. Fis. Acc. Lincei*, 2017, **28**, 19–27.
39 B. Kaewruksa, V. Vchirawongkwin and V. Ruangpornvisuti, *J. Mol. Struct.*, 2016, **1108**, 187–194.
40 F. Maleki and G. Pacchioni, *Top. Catal.*, 2020, **63**, 1717–1730.
41 M. Elstner, D. Porezag, G. Jungnickel, J. Elsner, M. Haugk, Th. Frauenheim, S. Suhai and G. Seifert, *Phys. Rev. B*, 1998, **58**, 7260–7268.
42 M. Elstner and G. Seifert, *Philos. Trans. R. Soc., A*, 2014, **372**, 20120483.
43 B. Hourahine, M. Berdakin, J. A. Bich, F. P. Bonafé, C. Camacho, Q. Cui, M. Y. Deshaye, G. Díaz Mirón, S. Ehlert, M. Elstner, T. Frauenheim, N. Goldman, R. A. González León, T. van der Heide, S. Irle, T. Kowalczyk, T. Kubař, I. S. Lee, C. R. Lien-Medrano, A. Maryewski, T. Melson, S. K. Min, T. Niehaus, A. M. N. Niklasson, A. Pecchia, K. Reuter, C. G. Sánchez, C. Scheurer, M. A. Sentef, P. V. Stishenko, V. Q. Vuong and B. Aradi, *J. Phys. Chem. A*, 2025, **129**, 5373–5390.
44 P. Sundarapura, S. Manzhos and M. Ihara, *Phys. Chem. Chem. Phys.*, 2023, **25**, 14566–14577.
45 D. Selli, G. Fazio, G. Seifert and C. Di Valentin, *J. Chem. Theory Comput.*, 2017, **13**, 3862–3873.
46 D. Selli, G. Fazio and C. Di Valentin, *J. Chem. Phys.*, 2017, **147**, 164701.
47 Q. Wang, M. Gu, C. Michel, N. Goldman, T. Niehaus and S. N. Steinmann, *J. Chem. Theory Comput.*, 2025, **21**, 5267–5278.
48 A. Penz, J. Gamper, J. M. Gallmetzer, F. R. S. Purtscher and T. S. Hofer, *J. Comput. Chem.*, 2025, **46**, e70140.

49 C. Panosetti, A. Engelmann, L. Nemec, K. Reuter and J. T. Margraf, *J. Chem. Theory Comput.*, 2020, **16**, 2181–2191.
50 M. Cui, K. Reuter and J. T. Margraf, *J. Chem. Theory Comput.*, 2024, **20**, 5276–5290.
51 P. Koskinen and V. Mäkinen, *Comput. Mater. Sci.*, 2009, **47**, 237–253.
52 J. J. Kranz, M. Kubillus, R. Ramakrishnan, O. A. von Lilienfeld and M. Elstner, *J. Chem. Theory Comput.*, 2018, **14**, 2341–2352.
53 G. Budiutama, R. Li, S. Manzhos and M. Ihara, *J. Chem. Theory Comput.*, 2023, **19**, 5189–5198.
54 P. Giannozzi, S. Baroni, N. Bonini, M. Calandra, R. Car, C. Cavazzoni, D. Ceresoli, G. L. Chiarotti, M. Cococcioni, I. Dabo, A. Dal Corso, S. de Gironcoli, S. Fabris, G. Fratesi, R. Gebauer, U. Gerstmann, C. Gougoussis, A. Kokalj, M. Lazzeri, L. Martin-Samos, N. Marzari, F. Mauri, R. Mazzarello, S. Paolini, A. Pasquarello, L. Paulatto, C. Sbraccia, S. Scandolo, G. Sclauzero, A. P. Seitsonen, A. Smogunov, P. Umari and R. M. Wentzcovitch, *J. Phys.: Condens. Matter*, 2009, **21**, 395502.
55 P. Giannozzi, O. Andreussi, T. Brumme, O. Bunau, M. Buongiorno Nardelli, M. Calandra, R. Car, C. Cavazzoni, D. Ceresoli, M. Cococcioni, N. Colonna, I. Carnimeo, A. Dal Corso, S. de Gironcoli, P. Delugas, R. A. DiStasio, A. Ferretti, A. Floris, G. Fratesi, G. Fugallo, R. Gebauer, U. Gerstmann, F. Giustino, T. Gorni, J. Jia, M. Kawamura, H.-Y. Ko, A. Kokalj, E. Küçükbenli, M. Lazzeri, M. Marsili, N. Marzari, F. Mauri, N. L. Nguyen, H.-V. Nguyen, A. Otero-de-la-Roza, L. Paulatto, S. Poncé, D. Rocca, R. Sabatini, B. Santra, M. Schlipf, A. P. Seitsonen, A. Smogunov, I. Timrov, T. Thonhauser, P. Umari, N. Vast, X. Wu and S. Baroni, *J. Phys.: Condens. Matter*, 2017, **29**, 465901.
56 D. Porezag, Th. Frauenheim, Th. Köhler, G. Seifert and R. Kaschner, *Phys. Rev. B*, 1995, **51**, 12947–12957.
57 B. Aradi, B. Hourahine and Th. Frauenheim, *J. Phys. Chem. A*, 2007, **111**, 5678–5684.
58 B. Hourahine, B. Aradi, V. Blum, F. Bonafé, A. Buccheri, C. Camacho, C. Cevallos, M. Y. Deshaye, T. Dumitrică, A. Dominguez, S. Ehlert, M. Elstner, T. van der Heide, J. Hermann, S. Irle, J. J. Kranz, C. Köhler, T. Kowalczyk, T. Kubař, I. S. Lee, V. Lutsker, R. J. Maurer, S. K. Min, I. Mitchell, C. Negre, T. A. Niehaus, A. M. N. Niklasson, A. J. Page, A. Pecchia, G. Penazzi, M. P. Persson, J. Řezáč, C. G. Sánchez, M. Sternberg, M. Stöhr, F. Stuckenberg, A. Tkatchenko, V. W. -z. Yu and T. Frauenheim, *J. Chem. Phys.*, 2020, **152**, 124101.
59 S. Fujii, K. Shimazaki and A. Kuwabara, *Acta Mater.*, 2024, **262**, 119460.
60 P. Virtanen, R. Gommers, T. E. Oliphant, M. Haberland, T. Reddy, D. Cournapeau, E. Burovski, P. Peterson, W. Weckesser, J. Bright, S. J. van der Walt, M. Brett, J. Wilson, K. J. Millman, N. Mayorov, A. R. J. Nelson, E. Jones, R. Kern, E. Larson, C. J. Carey, İ. Polat, Y. Feng, E. W. Moore, J. VanderPlas, D. Laxalde, J. Perktold, R. Cimrman, I. Henriksen, E. A. Quintero, C. R. Harris, A. M. Archibald, A. H. Ribeiro, F. Pedregosa and P. van Mulbregt, *Nat. Methods*, 2020, **17**, 261–272.
61 D. Dubbeldam, K. S. Walton, T. J. H. Vlugt and S. Calero, *Adv. Theory Simul.*, 2019, **2**, 1900135.
62 T. P. Senftle, S. Hong, M. M. Islam, S. B. Kylasa, Y. Zheng, Y. K. Shin, C. Junkermeier, R. Engel-Herbert, M. J. Janik, H. M. Aktulga, T. Verstraelen, A. Grama and A. C. T. van Duin, *npj Comput. Mater.*, 2016, **2**, 15011.
63 K. Kameda, T. Ariga, K. Ito, M. Ihara and S. Manzhos, *Digital Discovery*, 2024, **3**, 1967–1979.

64 M. Gaus, A. Goez and M. Elstner, *J. Chem. Theory Comput.*, 2012, **9**, 338–354.
65 G. Wulff, *Z. Kristallogr. - Cryst. Mater.*, 1901, **34**, 449–530.
66 A. Cadi-Essadek, A. Roldan and N. H. de Leeuw, *J. Phys. Chem. C*, 2015, **119**, 6581–6591.
67 Z. L. Wang and X. Feng, *J. Phys. Chem. B*, 2003, **107**, 13563–13566.
68 A. Bruix and K. M. Neyman, *Catal. Lett.*, 2016, **146**, 2053–2080.
69 A. S. Barnard and L. A. Curtiss, *Nano Lett.*, 2005, **5**, 1261–1266.
70 F. Zhang, Q. Jin and S.-W. Chan, *J. Appl. Phys.*, 2004, **95**, 4319–4326.
71 R. Grena, O. Masson, L. Portal, F. Rémondière, A. Berghout, J. Jouin and P. Thomas, *J. Phys. Chem. C*, 2015, **119**, 15618–15626.
72 A. Otero-de-la-Roza, M. A. Blanco, A. M. Pendás and V. Luaña, *Comput. Phys. Commun.*, 2009, **180**, 157–166.
73 A. Otero-de-la-Roza, E. R. Johnson and V. Luaña, *Comput. Phys. Commun.*, 2014, **185**, 1007–1018.
74 G. Henkelman, A. Arnaldsson and H. Jónsson, *Comput. Mater. Sci.*, 2006, **36**, 354–360.
75 E. Sanville, S. D. Kenny, R. Smith and G. Henkelman, *J. Comput. Chem.*, 2007, **28**, 899–908.
76 W. Tang, E. Sanville and G. Henkelman, *J. Phys.: Condens. Matter*, 2009, **21**, 084204.
77 E. Caldeweyher, C. Bannwarth and S. Grimme, *J. Chem. Phys.*, 2017, **147**, 034112.
78 E. Caldeweyher, S. Ehlert, A. Hansen, H. Neugebauer, S. Spicher, C. Bannwarth and S. Grimme, *J. Chem. Phys.*, 2019, **150**, 154122.
79 E. Caldeweyher, J.-M. Mewes, S. Ehlert and S. Grimme, *Phys. Chem. Chem. Phys.*, 2020, **22**, 8499–8512.
80 F. Legrain and S. Manzhos, *J. Power Sources*, 2015, **274**, 65–70.
81 D. Koch, M. Chaker, M. Ihara and S. Manzhos, *Molecules*, 2021, **26**, 5541.

# Supplementary Information

# Band Structure Modulation of $ZrO_2$ Nanoparticles for Control of CO Adsorption Properties: A Combined Density Functional Theory – Density Functional Tight Binding Study

Kexin Chen,[a] William Dawson,[b] Aulia Sukma Hutama,[c] Takahito Nakajima, [b] Keisuke Kameda,[a] Manabu Ihara,[a,1] Sergei Manzhos [a,2]

[a] School of Materials and Chemical Technology, Institute of Science Tokyo, Ookayama 2-12-1, Meguro-ku, Tokyo 152-8552, Japan

[b] RIKEN Center for Computational Science, Kobe, Hyogo 650-0047, Japan

[c] Department of Chemistry, Faculty of Mathematics and Natural Sciences, Universitas Gadjah Mada, Sekip Utara, Bulaksumur, Yogyakarta, 55281, Indonesia

[1] E-mail: mihara@chemeng.titech.ac.jp
[2] E-mail: manzhos.s.ss@m.titech.ac.jp

## S1 Zr-Zr/Zr–O $E_{Rep}$ parameterization details

The original DFTB3/3ob parameters were tested for CO adsorption on the $Zr_{85}O_{160}$ nanoparticle. The geometry optimization converged to 0.01 eV/ Å for the forces in more than 700 steps, and the resulting nanoparticle (NP) structure exhibited substantial distortion and appeared amorphous as shown in **Figure S1**. These results indicate that the original DFTB3/3ob parameters have limited transferability to CO adsorption on zirconia nanoparticles and are not directly suitable for describing such solid-state adsorption systems.

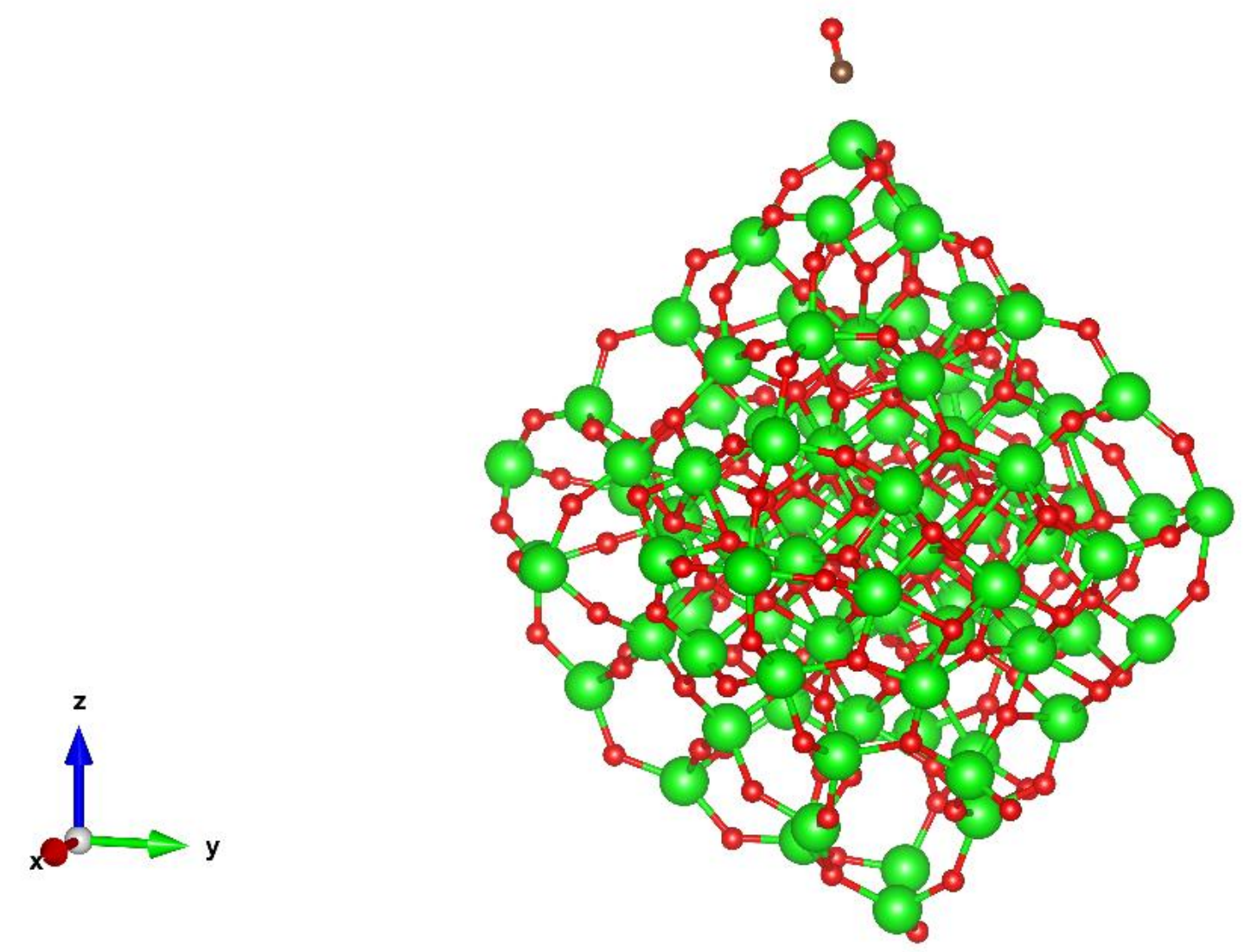

Figure S1. The final geometry of CO adsorption on the $Zr_{85}O_{160}$ nanoparticle computed with the original DFTB3/3ob parameters. Atom color scheme used here and throughout unless otherwise stated: Zr, green; O, red; C, brown.

The structural and energetic stability of DFTB parameters was evaluated by DFTB-based molecular dynamics (MD) annealing simulations for the $Zr_{85}O_{160}$ NP. The system was heated from 10 to 500 K over 2500 MD steps, equilibrated at 500 K for 2500 steps, and subsequently cooled to 100 K over 5000 steps using the Andersen thermostat with a time step of 1 fs.

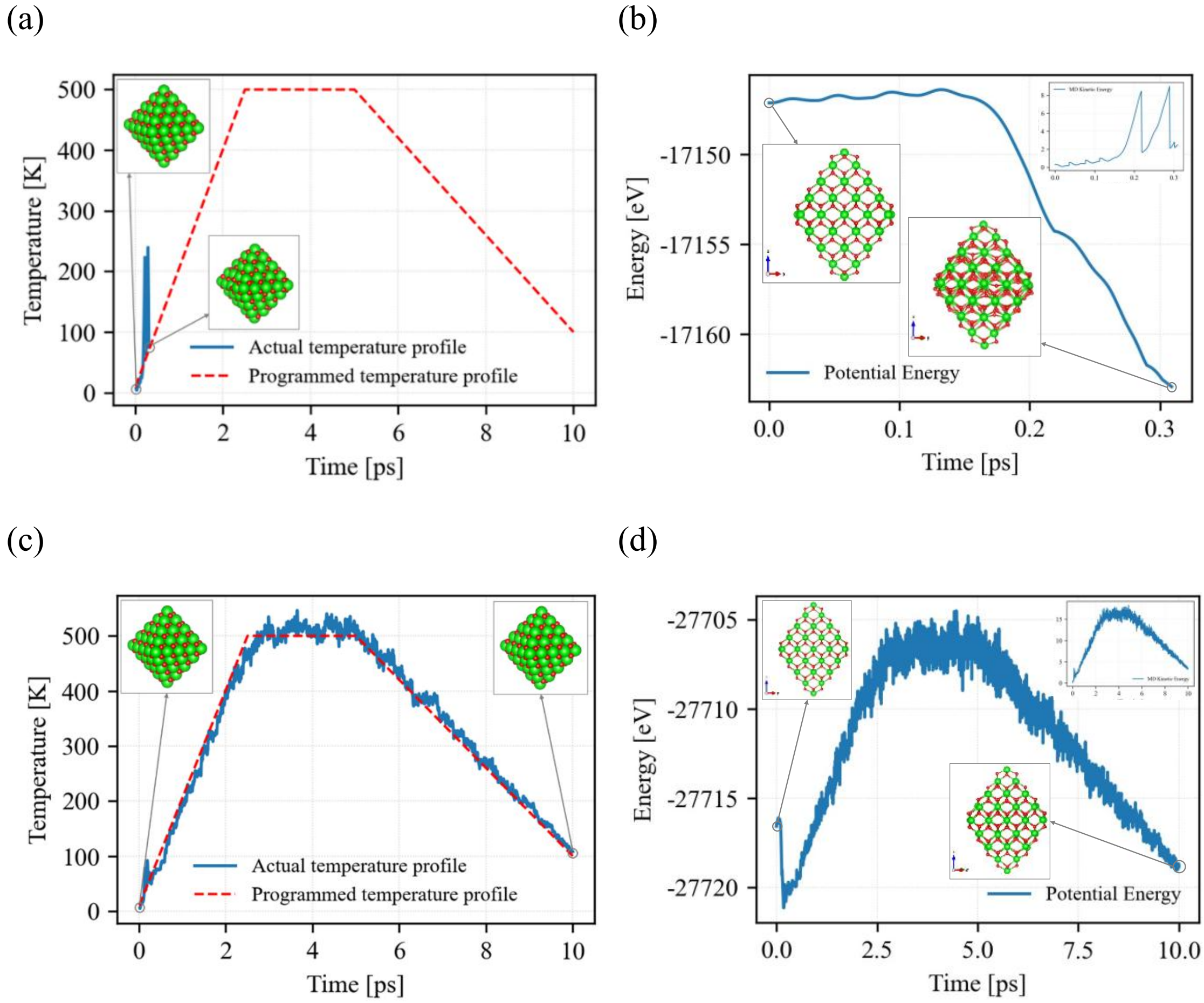

Figure S2. Molecular dynamics (MD) annealing simulations of the $Zr_{85}O_{160}$ nanoparticle: (a,c) temperature profiles and (b,d) potential energy evolution during annealing, with the corresponding MD kinetic energies shown in the insets. The upper panels (a,b) were calculated using the original DFTB3/3ob parameter set, while the lower panels (c,d) were calculated using DFTB–FF. The initial and final nanoparticle geometries are also shown as insets.

The MD temperature profiles and corresponding energy evolution obtained using the original DFTB3/3ob and DFTB–FF parameter sets are shown in **Figure S2**. With the original DFTB3/3ob parameters, the MD simulation progressed extremely slowly, completing only 309 of the prescribed 10,000 steps within two days using 96 MPI processes. In addition, the MD simulation using the original DFTB3/3ob parameters exhibited pronounced and irregular temperature fluctuations, with the instantaneous temperature deviating substantially from the prescribed annealing profile. These fluctuations were accompanied by a continuous decrease in potential energy and substantial structural reorganization, resulting in a loss of the original crystalline order. This behavior suggests that the nanoparticle tends to relax to a lower-potential-energy configuration and release energy, indicating that the original parameter set does not provide an adequate description of the structural energetics of the zirconia nanoparticle. After augmenting the DFTB repulsive potentials with classical long-range interatomic interactions, the temperature followed the prescribed annealing profile more closely, accompanied by physically reasonable evolution of the kinetic and potential energies. Moreover, the nanoparticle retained its overall structural integrity after MD annealing, without the nonphysical structural reorganization observed with the original DFTB3/3ob parameters.

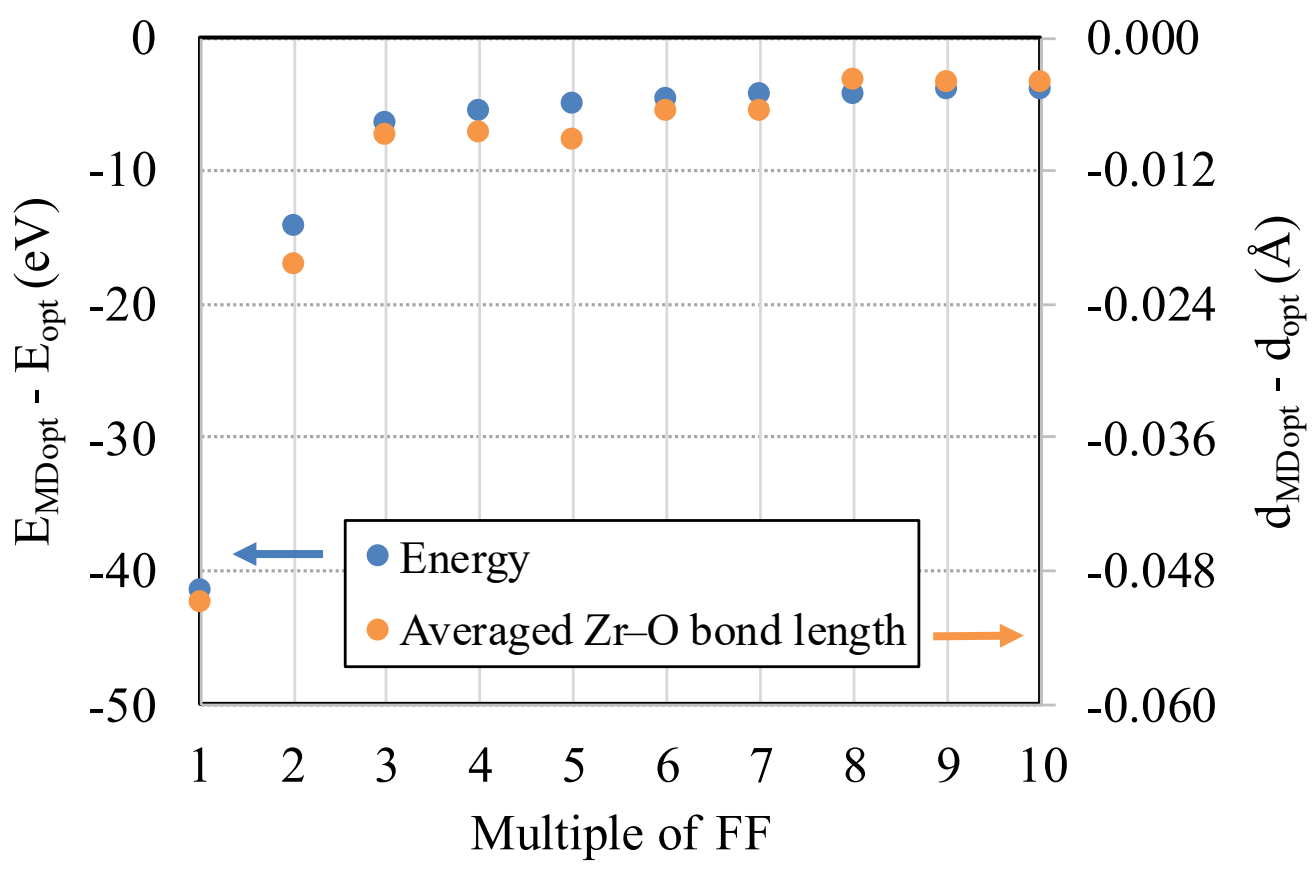

Figure S3. Energy and geometry stabilization effects of the DFTB–FF method with FF scaling factors ranging from 0 to 10.

To stabilize the NP structure toward that predicted by the FF, the FF was introduced into the DFTB parameters with a weighting factor. The hybrid DFTB/FF parameters, with the FF scaling factor ranging from 1 to 10, were tested and were found to yield pronounced improvements in both energetic and structural stability as shown in **Figure S3**. This work employs DFTB–FF parameters with a 4×FF contribution for all subsequent calculations, striking a balance between achieving structural stability and preserving the original DFTB framework to the greatest extent possible. The optimization results for the zirconia unit cells were benchmarked against the DFT reference results using the DFTB2/mio-based parameters,[1] the original DFTB3/3ob parameters, and the present DFTB–FF parameters with a 4×FF contribution, as summarized in **Table S1**. The results show improved accuracy of the DFTB–FF parameters and agreement comparable to those of the DFTB2 and DFT calculations. The original DFTB3/3ob repulsive potentials, the long-range FF potentials, and the total potential of DFTB–FF for Zr–Zr/Zr–O/O–O atomic pairs are shown in **Figure S4**.

Table S1. Bulk properties of zirconia unit cells optimized with DFT, DFTB2, DFTB3/3ob and DFTB–FF.

| Phase | Lattice parameters | DFT | DFTB2 | DFTB3/3ob | DFTB–FF |
|---|---|---|---|---|---|
| Cubic | $a$ (Å) | 5.0865 | 5.1317 | 5.0805 | 5.0852 |
| | $b$ (Å) | 5.0865 | 5.1317 | 5.0805 | 5.0852 |
| | $c$ (Å) | 5.0865 | 5.1317 | 5.0805 | 5.0852 |
| | $\alpha$ (°) | 90.0000 | 89.0324 | 89.9849 | 89.9940 |
| | $\beta$ (°) | 90.0000 | 89.0324 | 89.9849 | 89.9940 |
| | $\gamma$ (°) | 90.0000 | 89.0324 | 89.98499 | 89.9940 |
| Tetragonal | $a$ (Å) | 3.5981 | 3.5983 | 3.6544 | 3.5251 |
| | $b$ (Å) | 3.5981 | 3.5980 | 3.6441 | 3.6555 |
| | $c$ (Å) | 5.2082 | 5.2609 | 5.9465 | 5.2165 |
| | $\alpha$ (°) | 90.0000 | 90.0220 | 100.4378 | 90.0097 |
| | $\beta$ (°) | 90.0000 | 89.9949 | 79.5869 | 88.2576 |
| | $\gamma$ (°) | 90.0000 | 90.0199 | 90.0299 | 90.0000 |
| Monoclinic | $a$ (Å) | 5.1387 | 5.0985 | 5.2163 | 5.1855 |
| | $b$ (Å) | 5.2327 | 5.1247 | 4.5748 | 5.1300 |
| | $c$ (Å) | 5.3084 | 5.3575 | 6.1418 | 5.3631 |
| | $\alpha$ (°) | 90.0000 | 90.0010 | 90.2533 | 89.9990 |
| | $\beta$ (°) | 99.3026 | 100.7150 | 90.0057 | 99.0106 |
| | $\gamma$ (°) | 90.0000 | 89.9989 | 90.0142 | 89.9998 |

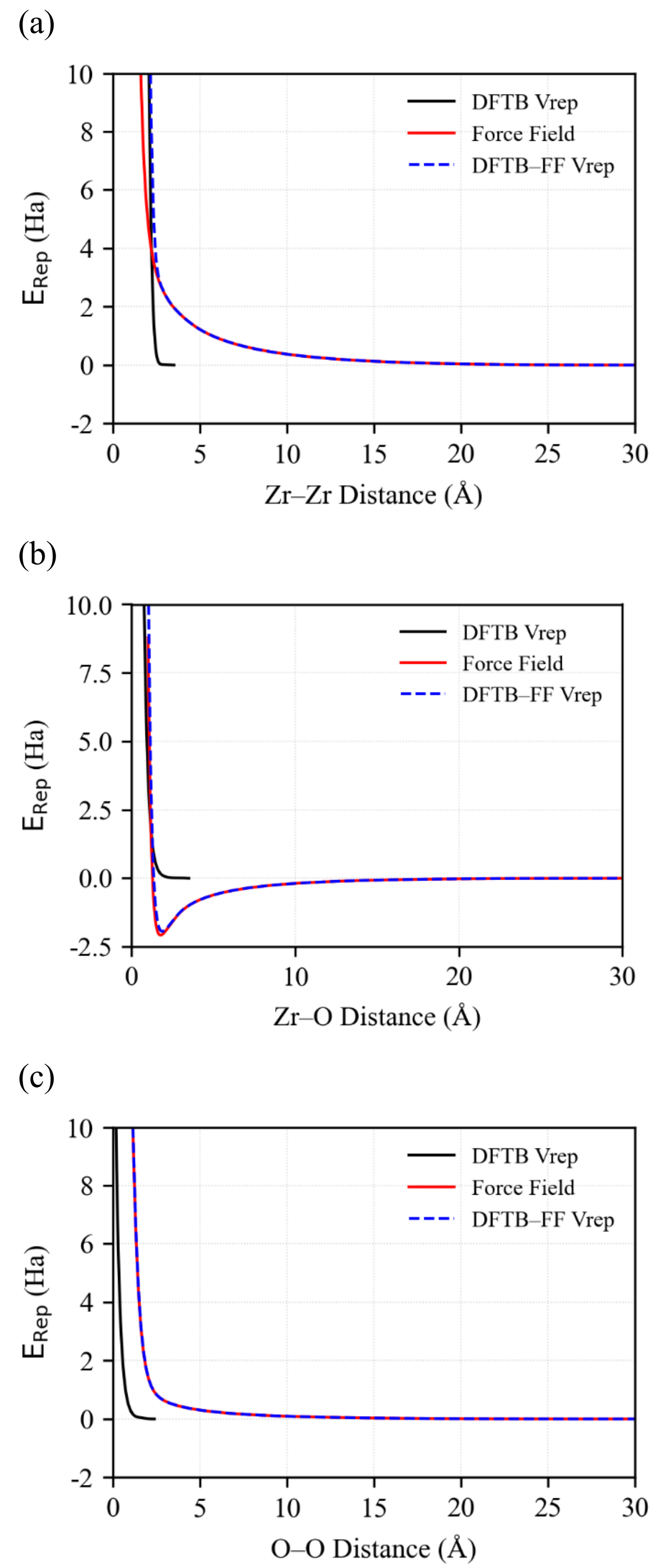

Figure S4. The original repulsive potentials, the long-range FF potentials, and resulting total interatomic potentials for the Zr–Zr/Zr–O/O–O atomic pairs in the hybrid DFTB–FF framework: (a) Zr–Zr, (b) Zr–O, and (c) O–O interactions.

## S2 Zr-C $E_{Rep}$ parameterization details

The variations in the local coordination environment and atomic charge shown in **Table S2**, together with the corresponding CO adsorption behavior (see **Table 2** in the main text for details), suggest that a single Zr–C repulsive potential may not adequately describe adsorption at chemically distinct surface sites. This motivates the use of coordination-dependent interatomic potentials to account for the different local environments of surface Zr atoms. Accordingly, two sets of Zr–C $E_{Rep}$ parameters were separately fitted to potential energy curves (PECs) calculated using DFT as reference data for bulk-like coordinated facet sites and under-coordinated tip/edge sites, respectively. **Figure S5** illustrates the representative Zr adsorption sites used for the two parameterizations, with bulk-like full-coordinated facet and under-coordinated tip/edge sites shown in green and blue, respectively. The resulting Zr–C interatomic potentials for the two coordination environments are presented with the original DFTB3/3ob potential in **Figure S6**.

Table S2. Atomic charges of Zr atoms at different sites in the *n*-type $Zr_{85}O_{160}$ nanoparticle calculated using DFT with Bader analysis and DFTB–FF with Mulliken analysis, with Bader charges derived from the DFTB electron density shown in parentheses. Bader charge analysis for DFTB–FF was performed by integrating the valence-electron density within the atomic basins defined by an all-electron density reconstructed by augmenting the valence-electron density with core-electron densities for all atoms in the system.

| System | Site | DFT Bader charge ($\|e\|$) | DFTB–FF charge ($\|e\|$) |
|---|---|---|---|
| $Zr_{85}O_{160}$ | $Zr_{Tip}$ | 1.7048 | 0.4611 (1.5173) |
| | $Zr_{Edge1}$ | 2.4482 | 0.5903 (1.9492) |
| | $Zr_{Edge2}$ | 2.5117 | 0.6140 (1.9993) |
| | $Zr_{Facet}$ | 2.5560 | 0.6147 (2.0576) |

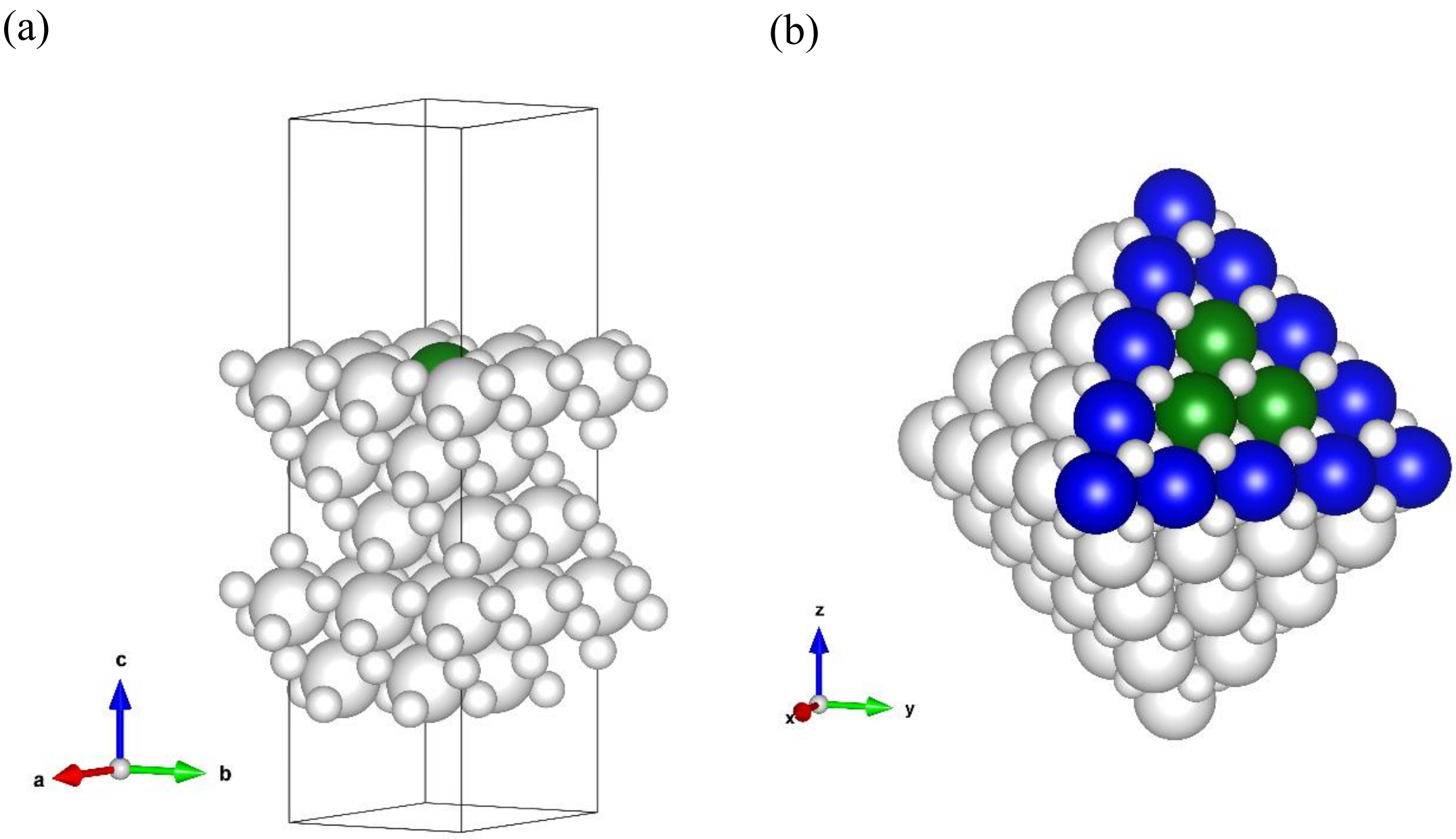

Figure S5. Reference systems for Zr–C interatomic potential fitting. (a) $ZrO_2$ (111) surface; (b) $Zr_{85}O_{160}$ nanoparticle. Facet and tip/edge Zr sites are represented in green and blue, respectively, whereas other Zr/O atoms not directly involved in the interatomic potential fitting are indicated in white.

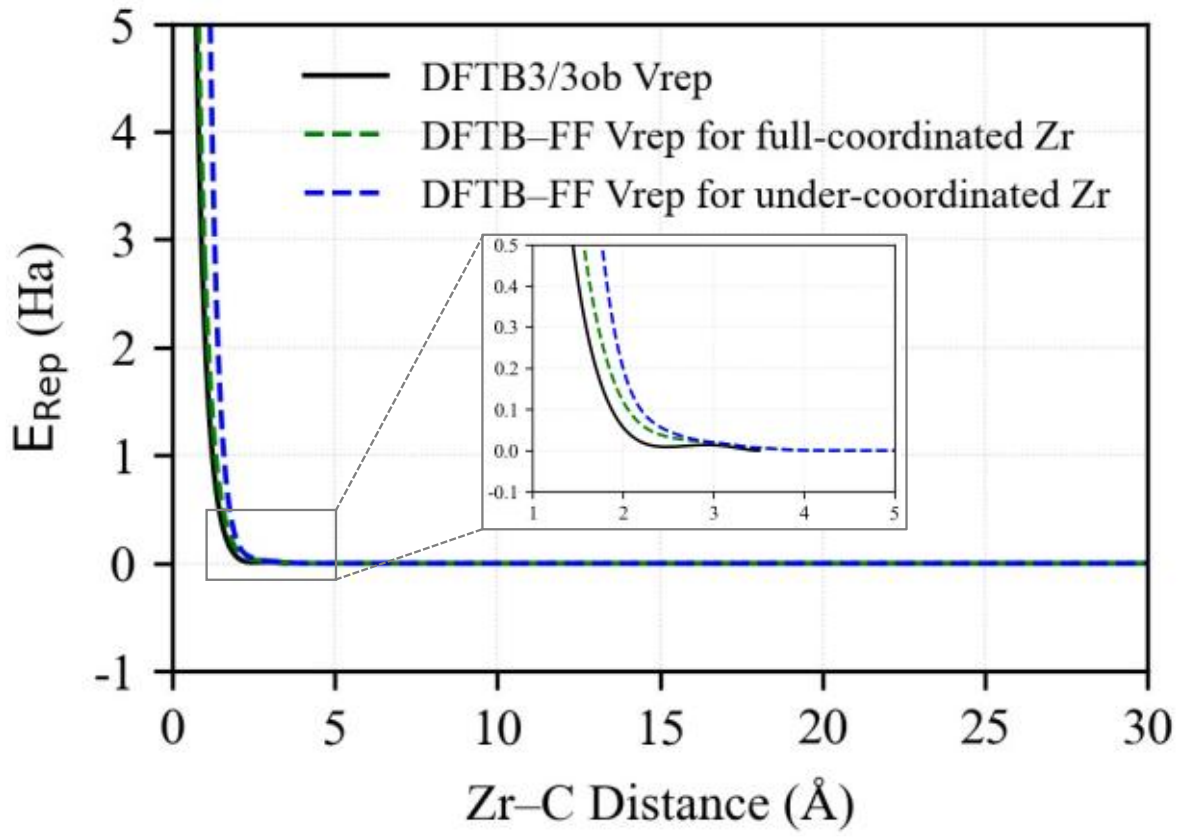

Figure S6. The interatomic repulsive potentials for Zr–C atomic pairs in the original DFTB3/3ob parameters plotted as a solid black line, and the re-parameterized ones for full-coordinated facet and under-coordinated tip/edge Zr sites plotted as green and blue dashed lines, respectively.

## S3 Supplementary calculation data

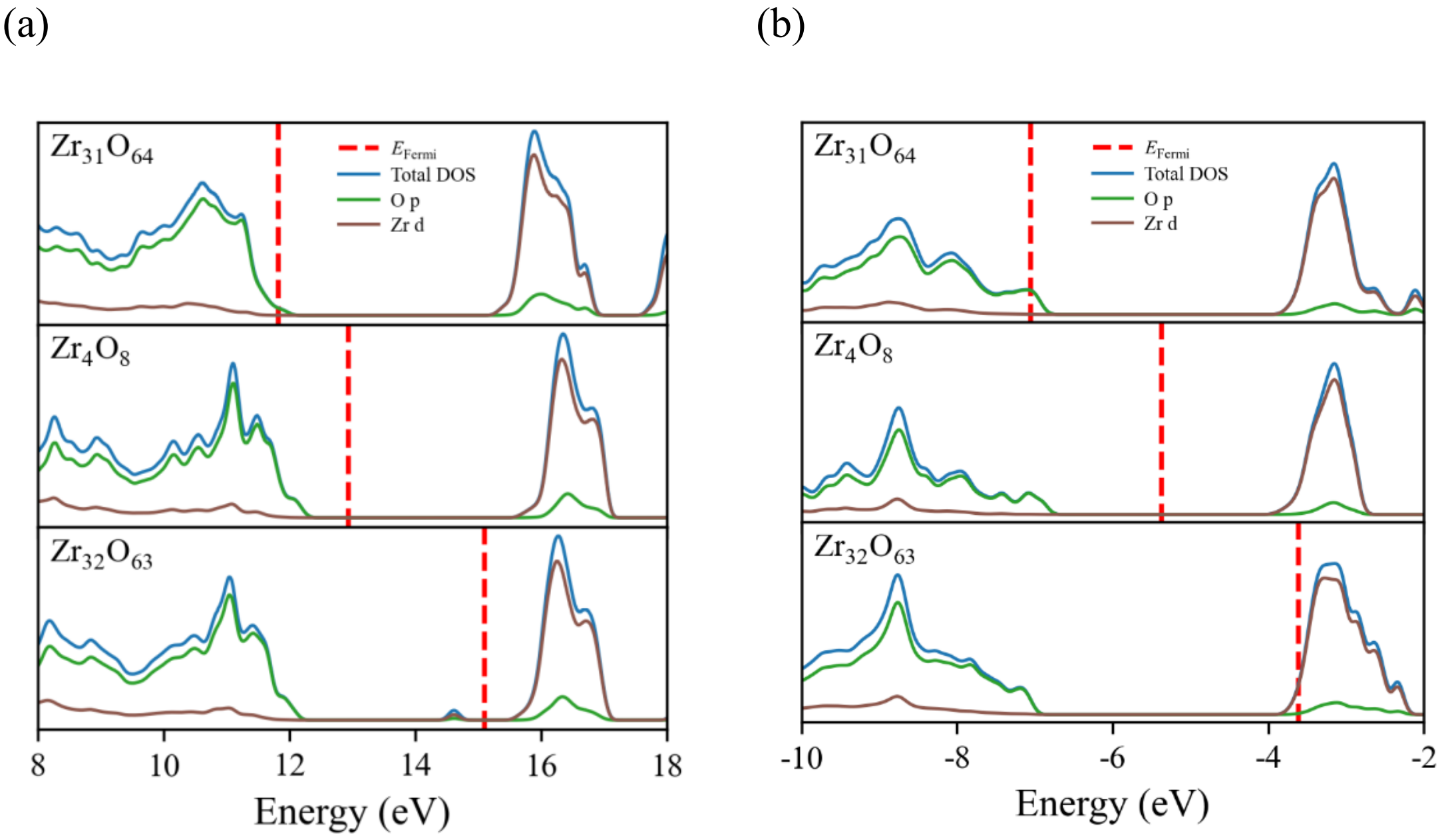

Figure S7. PDOS of cubic zirconia bulk of $Zr_{31}O_{64}$, $Zr_4O_8$, and $Zr_{32}O_{63}$ computed with (a) DFT, (b) DFTB–FF.

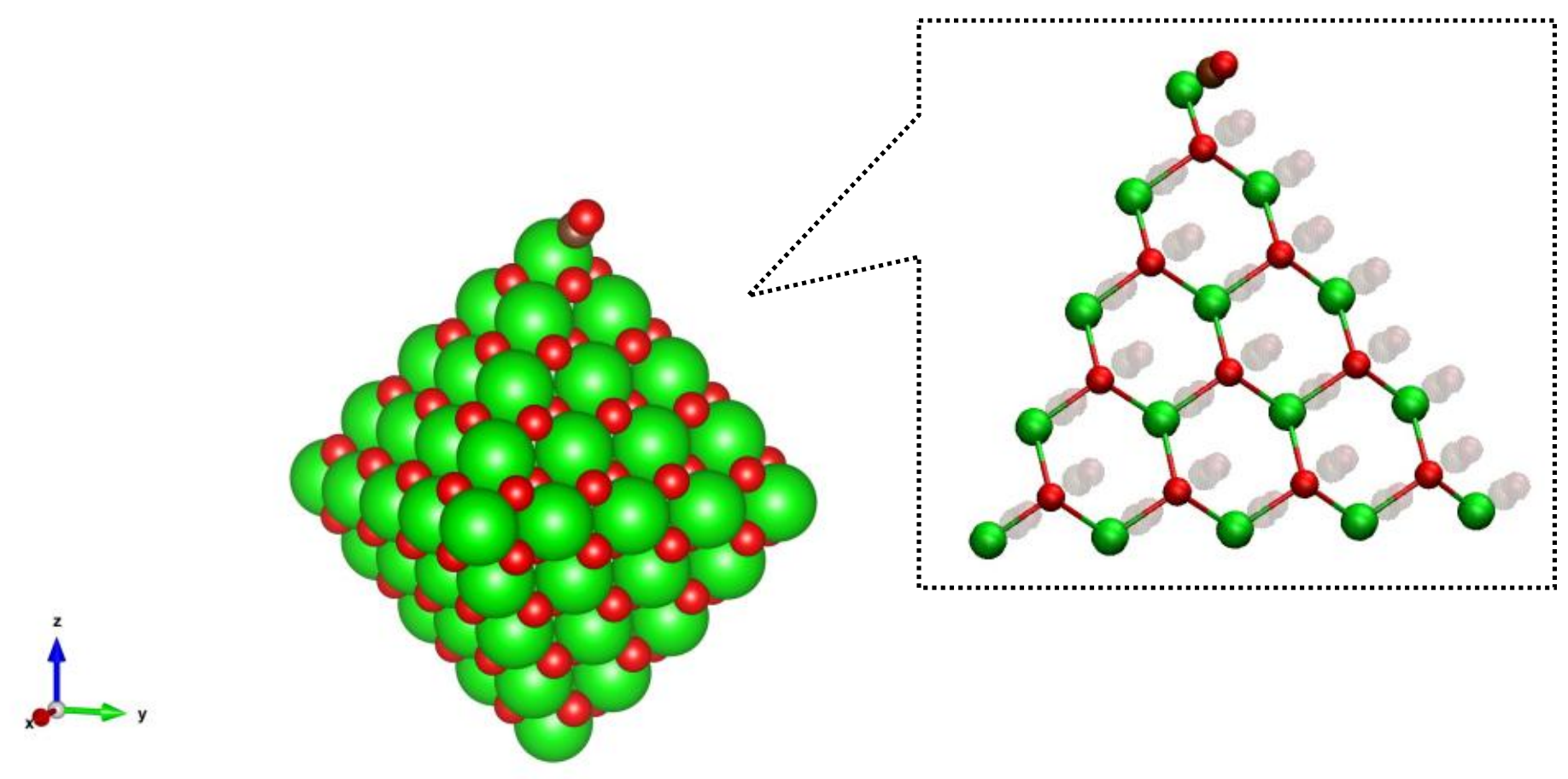

Figure S8. Initial CO adsorption configurations on the $Zr_{85}O_{160}$ nanoparticle surface.

The DFT-D4 model,[2–4] which provides a charge-dependent London dispersion correction, was applied to the DFT-optimized geometries to evaluate the contribution of dispersion interactions to the CO adsorption energies. The PBE functional was employed for the DFT-D4 calculations. As shown in **Table S3**, DFT-D4 calculations indicate a dispersion contribution of approximately 0.1 eV to the CO adsorption energy. As the present parameterization aims to reproduce PBE-level chemisorption energetics and the electronic-structure-based mechanism of CO bond weakening (not affected by dispersion corrections), the Zr–C repulsive potentials were fitted consistently to (non-dispersion-corrected) PBE reference data.

Table S3. CO adsorption energy ($E_{ad}$, eV) and corresponding dispersion correction energy ($E_{ad}^{DFT-D4}$, eV) at different Zr adsorption sites of $Zr_{79}O_{160}$, $Zr_{80}O_{160}$, and $Zr_{85}O_{160}$ nanoparticles computed with the DFT-D4 model.

| System | Type of band-structure | Adsorption site | $E_{ad}$ | $E_{ad}^{DFT-D4}$ |
|---|---|---|---|---|
| $Zr_{79}O_{160}$ | *p* | $Zr_{corner}$ | -0.47 | -0.10 |
| | | $Zr_{edge}$ | -0.43 | -0.11 |
| | | $Zr_{facet}$ | -0.28 | -0.13 |
| $Zr_{80}O_{160}$ | *Intrinsic* | $Zr_{tip}$ | -0.75 | -0.05 |
| | | $Zr_{edge1}$ | -0.31 | -0.09 |
| | | $Zr_{edge2}$ | -0.41 | -0.09 |
| | | $Zr_{edge3}$ | -0.43 | -0.10 |
| | | $Zr_{corner1}$ | -0.51 | -0.12 |
| | | $Zr_{corner2}$ | -0.49 | -0.10 |
| | | $Zr_{facet1}$ | -0.38 | -0.12 |
| | | $Zr_{facet2}$ | -0.26 | -0.12 |
| $Zr_{85}O_{160}$ | *n* | $Zr_{tip}$ | -0.60 | -0.08 |
| | | $Zr_{edge1}$ | -0.54 | -0.11 |
| | | $Zr_{edge2}$ | -0.63 | -0.11 |
| | | $Zr_{facet}$ | -0.25 | -0.12 |

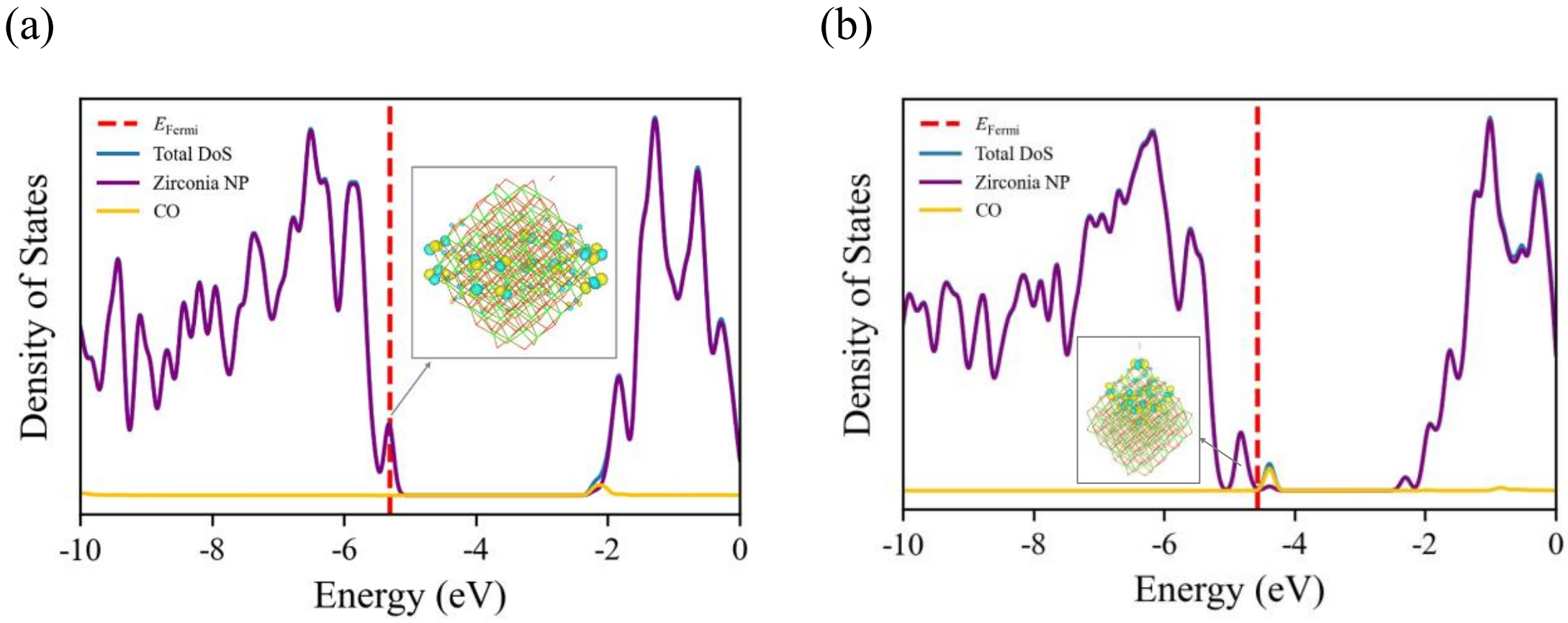

Figure S9. Densities of states computed with DFT for (a) CO adsorbed on $Zr_{79}O_{160}$, (b) CO adsorbed on $Zr_{80}O_{160}$. The molecular orbitals corresponding to the occupied orbitals near $E_{Fermi}$ are shown in the inset, while no LUMO-derived CO orbital was observed.

## S4 References

1 A. S. Hutama, L. A. Marlina, C.-P. Chou, S. Irle and T. S. Hofer, *ACS Omega*, 2021, **6**, 20530–20548.
2 E. Caldeweyher, C. Bannwarth and S. Grimme, *J. Chem. Phys.*, 2017, **147**, 034112.
3 E. Caldeweyher, S. Ehlert, A. Hansen, H. Neugebauer, S. Spicher, C. Bannwarth and S. Grimme, *J. Chem. Phys.*, 2019, **150**, 154122.
4 E. Caldeweyher, J.-M. Mewes, S. Ehlert and S. Grimme, *Phys. Chem. Chem. Phys.*, 2020, **22**, 8499–8512.